\documentclass[twocolumn,10pt]{iopjournal}
\usepackage{cpc-format}
\graphicspath{{figures/}}
\hypersetup{pdftitle={Enhanced structure-dependent alpha decay tunneling in deformed odd-odd nuclei via intense two-color laser pulses},
pdfauthor={Jun-Hao Cheng, Qiong Xiao, Liang Gao, Tong-Pu Yu},
pdfsubject={CPC-2026-0795: final clean manuscript}}

\begin{document}
\twocolumn[{
\begin{@twocolumnfalse}
\title{Enhanced structure-dependent $\alpha$ decay tunneling in deformed odd-odd nuclei via intense two-color laser pulses\textsuperscript{*}}
\author{Jun-Hao Cheng (\cn{程俊皓})$^{1}$, Qiong Xiao (\cn{肖琼})$^{2}$,
Liang Gao (\cn{高亮})$^{1,\dagger}$, and Tong-Pu Yu (\cn{余同普})$^{2,\ddagger}$}
\affil{$^1$Southwestern Institute of Physics, Chengdu 610003, People's Republic of China}
\affil{$^2$College of Science, National University of Defense Technology, Changsha 410073, People's Republic of China}
\begin{abstract}
In the present work, we investigate the effect of ultra-intense laser fields on the $\alpha$ decay tunneling of deformed odd-odd nuclei. Calculations performed within the deformed density-dependent cluster model show that, for deformed odd-odd $\alpha$ emitters, the laser-induced variation in the penetration probability separates into distinct branches associated with different valence particle-hole configurations relative to the $Z=82$ and $N=126$ shell closures. In contrast to the even-even and odd-$A$ cases, the odd-odd parent nucleus most sensitive to the laser field is not the one with the lowest $\alpha$ decay energy. This indicates that the laser response is governed not only by $Q_{\alpha}$ but also by structural correlations associated with the simultaneously unpaired proton and neutron. We further examine the pulse-averaged response under a two-color laser field. The results show that an optimized two-color field enhances the pulse-averaged laser response for most of the studied nuclei, while highly sensitive nuclei exhibit an effective gain equivalent to increasing the laser intensity by more than one order of magnitude. These findings identify structural determinants of laser-assisted $\alpha$ decay in odd-odd nuclei and provide guidance for selecting candidate nuclei and optimizing laser parameters in future experiments.
\end{abstract}
\vspace{1.2em}
\end{@twocolumnfalse}
}]
\begingroup
\renewcommand{\thefootnote}{\fnsymbol{footnote}}
\footnotetext[1]{This work was supported by the Innovation Program of Southwestern Institute of Physics (Grant No. 202501WDZCQN011), the Fundamental Research Program of China National Nuclear Corporation (Grant No. CNNC-JCKY-202632), the National Key R\&D Program of China (Grant Nos. 2023YFF0714905, 2024YFE03260100, and 2022YFE03200200) and the National Natural Science Foundation of China (Grant Nos. 1256050059 and 12375244) }
\footnotetext[2]{Corresponding author. E-mail: \href{mailto:gaoliang@swip.ac.cn}{gaoliang@swip.ac.cn}}
\footnotetext[3]{Corresponding author. E-mail: \href{mailto:tongpu@nudt.edu.cn}{tongpu@nudt.edu.cn}}
\endgroup
\setcounter{footnote}{0}

\section{Introduction}
\label{Sec.I}

Over the past two decades, the rapid development of radioactive ion beam facilities has continuously extended the boundary of the nuclear chart and has provided a rich platform for exploring exotic decay modes and nuclear structure far from stability \cite{doi:10.1146/annurev.nucl.55.090704.151604, RevModPhys.72.733, RevModPhys.84.567, PhysRevLett.110.242502, PhysRevC.87.044335, PhysRevC.91.051302, 10.1140/epja/i2015-15088-9, Xu2023}. Among the known decay modes, $\alpha$ decay occupies a unique position because it is both a sensitive probe of heavy and superheavy nuclear structure and a representative example of quantum tunneling through a potential barrier \cite{gamow1928quantentheorie, gurney1928wave, PhysRevLett.104.042701, Xiao20231, Deng20251, Deng20201, Deng20211}. At the same time, light has evolved from a spectroscopic probe into an external field with controllable intensity, frequency, and pulse duration. Owing to chirped-pulse amplification and large laser facilities, peak laser intensities have reached the order of $10^{23}\,\mathrm{W}/\mathrm{cm}^{2}$ in laboratories \cite{Yoon:21}, while the Extreme Light Infrastructure for Nuclear Physics (ELI-NP) \cite{Mi_icu_2019, doi:10.1063/1.5093535} and the Shanghai Superintense Ultrafast Laser Facility (SULF) \cite{Li:18, Yu:18} aim at even stronger fields. These advances make it timely to examine how an external electromagnetic field modifies nuclear tunneling and to identify which nuclear properties control the magnitude of laser-assisted decay effects. Moreover, experimental studies on femtosecond pumping of nuclear isomers and resonant X-ray excitation of nuclear states have further demonstrated that laser-nuclear interactions are becoming an experimentally accessible frontier \cite{PhysRevLett.128.052501, ko2023,Xiao20251,Xu20241,Zou20261,Zou20262, Yang2025, Qi2026,xiao2026advances}. Therefore, laser-assisted nuclear decay provides a valuable bridge between intense-light physics and low-energy nuclear structure.

Recent theoretical studies have shown that ultra-intense laser fields can change charged-particle decay half-lives by modifying the penetration probability through the electric-dipole interaction term \cite{Mi_icu_2013, PhysRevLett.119.202501, Kis_2018, BAI201823, PhysRevC.99.044610, PhysRevLett.124.212505, Xiao20241,Zou20241,Wang2025}. In our previous work on even-even nuclei, an analytical relation was derived to reveal the negative correlation between the $\alpha$ decay energy and the relative change in the penetration probability \cite{CHENG2024138322}. This was later extended to deformed odd-$A$ nuclei, where the laser response was found to be influenced by the shell effect and odd-even staggering \cite{cjh2025}. However, odd-odd nuclei have not yet been systematically investigated in this context. Unlike even-even nuclei, odd-odd nuclei contain both an unpaired proton and an unpaired neutron, and the residual proton-neutron correlation, possible particle-hole configurations, and nonzero angular-momentum transfer can introduce additional structural dependence in $\alpha$ decay. Previous studies of ordinary $\alpha$ decay have already shown that the odd-proton and/or odd-neutron degrees of freedom are commonly reflected through hindrance factors or structure-dependent formation probabilities \cite{CHENG2019350, Wang:2022axn}. Thus, odd-odd nuclei provide an important testing ground for determining whether the laser-induced change in half-life is governed only by the decay energy or is also modulated by the coupled unpaired proton-neutron structure.

In the present work, we systematically investigate laser-assisted deformed $\alpha$ decay of ground-state odd-odd nuclei under a state-of-the-art laser pulse with a peak intensity of $10^{23}\,\mathrm{W}/\mathrm{cm}^{2}$. The $\alpha$ decay energies, half-lives, spin-parity information, and ground-state deformation parameters are taken from the latest evaluated nuclear data and deformation tables \cite{NUBASE2020, CPC-2021-0034, CPC-2020-0033, MOLLER20161}. Within the deformed density-dependent cluster model and the WKB approximation, we calculate the relative change in the penetration probability and the corresponding relative change in the $\alpha$ decay half-life for the selected odd-odd nuclei. Particular attention is paid to the influence of the simultaneously unpaired proton and neutron on the laser response. In addition to identifying the nuclei most sensitive to the external laser field, we analyze how the $\Delta P$--$Q_{\alpha}$ relationship is modified by odd-odd nuclear structure and how the valence proton-neutron configuration around major shell closures, especially near $Z=82$ and $N=126$, affects the systematic behavior of the laser-induced half-life variation. This allows the present study to go beyond a direct extension of the even-even and odd-$A$ calculations and to clarify the additional structural information carried by odd-odd nuclei in laser-assisted $\alpha$ decay. Furthermore, because a real measurement is performed with a finite laser pulse rather than a static field, we also examine the complete-pulse average response and introduce a two-color pulse to reduce the cancellation caused by the oscillating electric field. This connects the selection of laser-sensitive odd-odd nuclei with the optimization of experimentally accessible laser-pulse configurations.

This article is organized as follows. In the next section, the theoretical framework for calculating the $\alpha$ decay half-life and the laser-nucleus interaction in deformed odd-odd nuclei is described in detail. In Section \ref{Sec.III}, the numerical results are presented, including the calculated relative changes in the penetration probability and half-life, the systematics of the laser response, the role of unpaired proton-neutron effects, and the enhancement of the complete-pulse average response using a two-color pulse. The final section provides a brief summary and outlook.

\section{Theoretical framework}\label{Sec.II}

\begin{figure*}[t]
\centering
\includegraphics[width=0.76\textwidth]{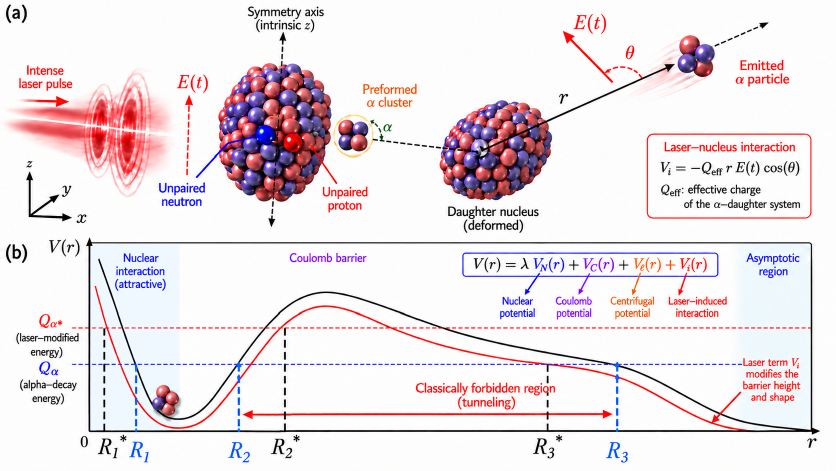}
\caption{Schematic illustration of laser-assisted deformed $\alpha$ decay. (a) Geometry of the process for a deformed odd-odd parent nucleus. The unpaired neutron and unpaired proton indicate the additional structure of the odd-odd system, and the laser electric field couples to the relative coordinate between the emitted $\alpha$ particle and the deformed daughter nucleus. (b) Barrier picture used in the WKB treatment. The total potential contains the nuclear, Coulomb, centrifugal, and laser-induced interaction terms. The blue dashed line denotes the ordinary decay energy $Q_{\alpha}$, whereas the red dashed line denotes the laser-modified energy $Q_{\alpha}^{*}$. The corresponding turning points $R_i$ and $R_i^{*}$ illustrate how the laser term changes the tunneling region.}
\label{fig1}
\end{figure*}

Before presenting the detailed formulas, the physical picture of laser-assisted deformed $\alpha$ decay is summarized in Fig. \ref{fig1}. The preformed $\alpha$ cluster is emitted from a deformed odd-odd parent nucleus, and the laser electric field acts on the relative coordinate between the emitted $\alpha$ particle and the daughter nucleus through the effective charge. In this picture, the nuclear, Coulomb, and centrifugal potentials form the static barrier, while the laser-induced interaction modifies the barrier height and shape. The notation shown in Fig. \ref{fig1} therefore provides a direct guide for the orientation averaging, the laser-field projection, and the classical turning points used in the WKB calculation below.

\subsection{The theoretical method}

The $\alpha$ decay half-life $T_{1/2}$, an important indicator of nuclear stability, is written as
\begin{equation}
\label{eq1}
T_{1/2}=\frac{\hbar \ln 2}{\Gamma}, 
\end{equation}
where $\hbar$ is the reduced Planck constant and $\Gamma$ is the $\alpha$ decay width, written as \cite{PhysRevC.73.041301}
\begin{equation}
\label{eq2}
\Gamma=\frac{\hbar^{2}}{4\mu}S_{\alpha}FP, 
\end{equation}
where $S_{\alpha}$, $F$, and $P$ are the $\alpha$-particle formation probability, the normalization factor, and the penetration probability, respectively. $\mu=\frac{M_d M_{\alpha}}{M_d+M_{\alpha}}$ is the reduced mass of the $\alpha$ particle and the daughter nucleus in the center-of-mass coordinate. Here, $M_{d}$ and $M_\alpha$ are the masses of the daughter nucleus and the $\alpha$ particle, respectively.

To include nuclear deformation, the total penetration probability $P$ is obtained by averaging $P_{\varphi}$ over all orientations. This method is frequently used in calculations of $\alpha$ decay and fusion reactions \cite{PhysRevC.85.044324, PhysRevC.86.044317, NI2015108, PhysRevC.83.044317, STEWART1996332} and is expressed as \cite{PhysRevC.73.041301}
\begin{equation}
P=\frac{1}{2}\int_{0}^{\pi}P_{\varphi}\, \sin\varphi\, d\varphi, 
\label{eq4}
\end{equation}
\begin{equation}
P_{\varphi}=\exp\! [- 2 \int_{R_{2}}^{R_{3}} k(r, t, \varphi, \theta)\, dr].
\label{eq5}
\end{equation}
where $r$ is the separation between the centers of mass of the $\alpha$ particle and the daughter nucleus. The variable $\varphi$ denotes the orientation angle of the symmetry axis of the daughter nucleus relative to the emitted $\alpha$ particle. Meanwhile, $\theta$ is related to the laser-nucleus interaction, which will be explained in more detail in the following subsection. The classical turning points, denoted by $R_{1}$, $R_{2}$, and $R_{3}$, are calculated from $V(r, t, \varphi, \theta) = Q_{\alpha}$. The wavenumber $k(r, t, \varphi, \theta)$ is expressed as
\begin{equation}
k(r, t, \varphi, \theta)= \sqrt{\frac{2\mu}{\hbar^{2}} \left|V(r, t, \varphi, \theta)-Q_{\alpha}\right| }, 
\end{equation}
where $Q_{\alpha}$ is the $\alpha$ decay energy. Similarly, the normalization factor $F$ is obtained from
\begin{equation}
F=\frac{1}{2}\int_{0}^{\pi} F_{\varphi}\, \sin\varphi\, d\varphi, 
\end{equation}
\begin{equation}
F_{\varphi}= \frac{1}{\int_{R_{1}}^{R_{2}} \frac{1}{2 k(r,t,\varphi,\theta)}\, dr},
\end{equation}

In this work, the total interaction potential $V(r, t, \varphi, \theta)$ between the daughter nucleus and the emitted $\alpha$ particle is written as
\begin{equation}
\begin{aligned}
V(r, t, \varphi, \theta)={}&\lambda(\varphi) V_N(r, \varphi)+V_l(r)\\
&+V_C(r, \varphi)+V_i(r, t, \varphi, \theta).
\end{aligned}
\label{eq8}
\end{equation}
where $V_l(r)$, $V_C(r, \varphi)$, and $V_N(r, \varphi)$ are the centrifugal, Coulomb, and nuclear potentials, respectively. The interaction between the electromagnetic field and the decay system is described by $V_i(r, t, \varphi, \theta)$ \cite{Mi_icu_2013}, and a detailed explanation of this interaction is provided in the following subsection. The value of $\lambda(\varphi)$ can be determined using the Bohr-Sommerfeld quantization condition.

The deformed Coulomb potential is obtained using the double-folding model, written as \cite{PhysRevC.73.041301}
\begin{equation}
V_C(\vec{r}, \varphi)=\int\!\int\frac{\rho_{d}(\vec{r}_d)\rho_{\alpha}(\vec{r}_\alpha)}{|\vec{r}+\vec{r}_d-\vec{r}_\alpha|}\, d\vec{r}_\alpha\, d\vec{r}_d, 
\label{eq:coulomb-folding}
\end{equation}
where $\vec{r}_d$ and $\vec{r}_\alpha$ represent the radius vectors in the charge distributions of the daughter nucleus and the emitted $\alpha$ particle, respectively. $\rho_{d}$ and $\rho_{\alpha}$ represent the density distributions of the daughter nucleus and the emitted $\alpha$ particle, respectively. Nuclear densities and the corresponding interaction potentials can be described using different theoretical approaches, such as the microscopic deformed relativistic Hartree-Bogoliubov theory in continuum (DRHBc) \cite{Choi2024DRHBcAlpha,Zhang2025DRHBcReview} and its triaxial extension (TRHBc) \cite{Lu2024TRHBcProton,Zhang2025TRHBc}, among others. In the present calculation, we adopt the analytical Coulomb potential obtained by Fourier transformation for uniform charge distributions with sharp surfaces \cite{PhysRevC.61.044607, ISMAIL200353, Gao_Long_2008, Gui2022DeformedTPA}. Following our previous treatment \cite{Xiao20241}, the Coulomb potential is approximated as  
\begin{equation}
V_C(\vec{r}, \varphi)=V_C^{(0)}(\vec{r}, \varphi)+V_C^{(1)}(\vec{r}, \varphi)+V_C^{(2)}(\vec{r}, \varphi), 
\label{eq:coulomb-expanded}
\end{equation}
where $V_C^{(0)}(\vec{r}, \varphi)$, $V_C^{(1)}(\vec{r}, \varphi)$, and $V_C^{(2)}(\vec{r}, \varphi)$ are the bare Coulomb interaction, linear Coulomb coupling, and second-order Coulomb coupling, respectively \cite{PhysRevC.61.044607}. In this approximation, the charge density dependence is incorporated into the nuclear charges, radii, and deformation parameters. Thus, evaluating Eq.~(\ref{eq:coulomb-expanded}) requires no separate specification or numerical integration of $\rho_d$ and $\rho_\alpha$.

The Langer-modified $V_l(r)$ is taken in the form \cite{doi:10.1063/1.531270}
\begin{equation}
V_{l}(r)=\frac{\hbar^2(l+\frac{1}{2})^2}{2{\mu}r^2}, 
\label{eq:centrifugal}
\end{equation}

\begingroup
the orbital angular momentum $l$ carried by the $\alpha$ particle is taken as the minimum allowed value $l_{\min}$. Based on the conservation of angular momentum and parity, $l_{\min}$ is determined by \cite{CHENG2019350}
\begin{equation}
l_{\min}=\begin{cases}
\Delta_j, & \text{for even }\Delta_j\text{ and }\pi_p=\pi_d,\\
\Delta_j+1, & \text{for even }\Delta_j\text{ and }\pi_p\ne\pi_d,\\
\Delta_j, & \text{for odd }\Delta_j\text{ and }\pi_p\ne\pi_d,\\
\Delta_j+1, & \text{for odd }\Delta_j\text{ and }\pi_p=\pi_d,
\end{cases}
\label{eq:minimum-angular-momentum}
\end{equation}
where $\Delta_j=|j_p-j_d|$. Here, $j_p$ ($j_d$) and $\pi_p$ ($\pi_d$) denote the spin and parity of the parent (daughter) nucleus, respectively. The ground-state spin and parity assignments of the parent and daughter nuclei are taken from NUBASE2020 \cite{NUBASE2020}. The adopted $l$ values are listed in Table~\ref{tab1}.
\par\endgroup

In the present work, the nuclear potential $V_N(r, \varphi)$ between the emitted $\alpha$ particle and the daughter nucleus is described by the standard Woods--Saxon form \cite{PhysRevC.83.067302,BaiRen2018WSG,Wan2024WS}. 
\begingroup
\begin{equation}
V_N(r,\varphi)=\frac{V'}{1+\exp\{[r-R_d(\varphi)]/a(I_d)\}},
\end{equation}
where
\begin{equation}
\begin{aligned}
R_d(\varphi)={}&r_0(I_d)A_d^{1/3}
[1+\beta_2Y_{20}(\varphi)+\beta_4Y_{40}(\varphi)+\beta_6Y_{60}(\varphi)],\\
r_0(I_d)={}&r_{00}+r_{01}I_d,\qquad
a(I_d)=a_{00}+a_{01}I_d,
\end{aligned}
\end{equation}
and $I_d=(N_d-Z_d)/A_d$ is the isospin asymmetry of the daughter nucleus. The ground-state nuclear deformations are denoted by $\beta_2$, $\beta_4$, and $\beta_6$, representing quadrupole, hexadecapole, and hexacontatetrapole deformations, respectively. $A_d$, $N_d$, and $Z_d$ are the mass, neutron, and proton numbers of the daughter nucleus, and $Y_{\lambda0}(\varphi)$ is the spherical harmonic for $\lambda=2,4,6$. The globally fitted coefficients are
\begin{equation}
\begin{aligned}
V'&=-215.085\,\mathrm{MeV},\\
r_{00}&=0.886\,\mathrm{fm},& r_{01}&=-0.432\,\mathrm{fm},\\
a_{00}&=1.081\,\mathrm{fm},& a_{01}&=0.524\,\mathrm{fm}.
\end{aligned}
\label{eq:odd-odd-potential-parameters}
\end{equation}
The five coefficients are common to all investigated nuclei, whereas the actual nuclear potential varies with the daughter-nucleus isospin, mass, and deformation through $r_0(I_d)$, $a(I_d)$, and $R_d(\varphi)$, and with the strength factor $\lambda(\varphi)$ determined for each nucleus and orientation by the Bohr--Sommerfeld quantization condition. Here, the five coefficients are fitted to the selected odd-odd nuclei as a whole, without separate half-life fits for individual nuclei. Separate fits could further reduce the differences between calculated and experimental half-lives, but would make comparisons between nuclei depend on parameter adjustments specific to each measured half-life. This may weaken the test of common decay systematics. The common parametrization adopted here describes the decay systematics with a limited number of shared coefficients and provides a consistent basis for comparing the laser response.
\par\endgroup

\subsection{Laser-nucleus interaction}
\subsubsection{The quasistatic approximation}
The full width at half maximum (FWHM) of laboratory laser pulses with peak intensities exceeding $10^{23}\,\mathrm{W}/\mathrm{cm}^{2}$ is approximately $19.6\,\mathrm{fs}$ ($=1.96 \times 10^{-14}$ s) \cite{Yoon:21}. The optical periods of a near-infrared laser with a wavelength of approximately 800 nm and an X-ray free-electron laser \cite{9760631} with a photon energy of 10 keV are approximately $10^{-15}$ s and $10^{-19}$ s, respectively. For $\alpha$ decay, the emitted $\alpha$ particle oscillates back and forth at high frequencies inside the parent nucleus, with a small probability of tunneling out whenever the preformed $\alpha$ particle reaches the potential wall. Since the typical decay energy for $\alpha$ decay is several MeV, the velocity of the preformed $\alpha$ particle is approximately $10^7$ m/s and the size of the parent nucleus is approximately 1 fm. The oscillation frequency can therefore be roughly estimated as $10^{22}$ Hz. The length of the tunneling path is less than 100 fm, and the time required for the emitted $\alpha$ particle to pass through the barrier is less than $10^{-20}$ s. The optical period of the highest-intensity laser pulse currently achievable is much longer than this time. Therefore, the laser field does not change significantly during the passage of the emitted $\alpha$ particle through the potential barrier, and the process can be considered quasistatic. A similar quasistatic approximation is usually used to describe the tunneling ionization of atoms in strong-field atomic physics \cite{PhysRevA.54.R2551, PhysRevA.63.011404}. 

In addition, the kinetic energy of the emitted $\alpha$ particle is only a few MeV, so its velocity is much smaller than the speed of light in vacuum. This means that the effect of the laser electric field on the emitted $\alpha$ particle is expected to be much larger than that of the laser magnetic field. Therefore, the magnetic component of the laser field is neglected in the present work.

\subsubsection{Relative motion of daughter nuclei and \texorpdfstring{$\alpha$}{alpha} particles in laser fields in center-of-mass coordinates}

In the framework of the quasistatic approximation, the interaction between the daughter nucleus and the emitted $\alpha$ particle can be effectively described using the time-dependent Schr\"odinger equation (TDSE) \cite{PhysRevC.102.064629}, written as
\begin{equation}
i \hbar \frac{\partial \Phi(\vec{r}_\alpha, \vec{r}_d, t)}{\partial t}=H(t)\Phi(\vec{r}_\alpha, \vec{r}_d, t), 
\end{equation}
where $H(t)$ represents the time-dependent minimal-coupling Hamiltonian, which can be written as
\begin{equation}
H(t)=\sum_i\frac{1}{2m_i}[\vec{p}_i-\frac{q_i}{c}\vec{A}(t)]^2+V(r), 
\end{equation}
where the index $i$ denotes the daughter nucleus and the emitted $\alpha$ particle.

For the center-of-mass coordinates ($\vec{R}, \vec{P}, \vec{r}, \vec{p}$):
\begin{equation}
\begin{aligned}
&\ \vec{r}_\alpha=\vec{R}+m_d\vec{r}/(m_\alpha+m_d)\\
&\ \vec{p}_\alpha=\vec{p}+m_\alpha\vec{P}/(m_\alpha+m_d)\\
&\ \vec{r}_d=\vec{R}-m_\alpha \vec{r}/(m_\alpha+m_d)\\
&\ \vec{p}_d=-\vec{p}+m_d\vec{P}/(m_\alpha+m_d).\\
\end{aligned}
\end{equation}
The time-dependent minimal-coupling Hamiltonian is written as
\begin{equation}
H(t)=\frac{1}{2M}[\vec{P}-\frac{q}{c}\vec{A}(t)]^2+\frac{1}{2\mu}[\vec{p}-\frac{Q_{\rm eff}}{c}\vec{A}(t)]^2+V(r), 
\end{equation}
where $q=q_\alpha+q_d$ and $M=m_\alpha+m_d$. $Q_{\rm eff}$ represents the effective charge for relative motion, which describes the tendency of the laser electric field to separate the emitted $\alpha$ particle from the daughter nucleus. It is given by
\begin{equation}
Q_{\rm eff}=\frac{q_\alpha m_d-q_d m_\alpha}{M}.
\end{equation}

By introducing unitary transformations, the wave function can be transformed into the center-of-mass coordinates
\begin{equation}
\phi(\vec{r}, \vec{R}, t)=\hat{U}_r\hat{U}_R\Phi(\vec{r}, \vec{R}, t), 
\end{equation}
\begingroup\RaggedRight where $\hat{U}_r=\exp\left[-i\frac{Q_{\rm eff}}{c}\vec{A}(t)\cdot\vec{r}/\hbar\right]$ and $\hat{U}_R=\exp\left[-i\frac{q}{c}\vec{A}(t)\cdot\vec{R}/\hbar\right]$. The TDSE can be rewritten as\par\endgroup
\begin{eqnarray}
i \hbar \frac{\partial \phi(\vec{r}, \vec{R}, t)}{\partial t}=&[-\frac{{\hbar}^2}{2\mu}\nabla^2_r+V(r)-Q_{\rm eff}\vec{r}\cdot\vec{E}(t)\nonumber\\
&\ -\frac{{\hbar}^2}{2M}\nabla^2_R-q\vec{R}\cdot\vec{E}(t)]\phi(\vec{r}, \vec{R}, t), 
\end{eqnarray}
where $c\vec{E}(t)=-d\vec{A}(t)/dt$ is the time-dependent laser electric field. For simplicity, we factorize the wave function $\phi(\vec{r}, \vec{R}, t)$ into two parts: $\phi(\vec{r}, \vec{R}, t)=\chi_1(\vec{R}, t)\chi_2(\vec{r} , t)$. This allows the TDSE to be separated into two equations that describe the center-of-mass motion and the relative motion between the daughter nucleus and the emitted $\alpha$ particle. They are written as

\begin{equation}
i \hbar \frac{\partial\chi_1(\vec{R} , t)}{\partial t}=[-\frac{{\hbar}^2}{2M}\nabla^2_R-q\vec{R}\cdot\vec{E}(t)]\chi_1(\vec{R} , t), 
\end{equation}
\begin{equation}
i \hbar \frac{\partial\chi_2(\vec{r} , t)}{\partial t}=[-\frac{{\hbar}^2}{2\mu}\nabla^2_r+V(r)-Q_{\rm eff}\vec{r}\cdot\vec{E}(t)]\chi_2(\vec{r} , t).
\end{equation}

The equation describing the relative motion is directly coupled to the laser electric field. This coupling gives the interaction potential energy between the relative coordinate and the laser field,
\begin{equation}
V_i(\vec{r}, t, \theta)=-Q_{\rm eff}\vec{r}\cdot\vec{E}(t)=-Q_{\rm eff} r E(t)\cos\theta, 
\end{equation}
where $\theta$ is the angle between the vectors $\vec{r}$ and $\vec{E}(t)$.

\begin{table*}[!htbp]
\centering
\scriptsize
\setlength{\tabcolsep}{1.3pt}
\renewcommand{\arraystretch}{1.4}
\caption{Comparison of experimental and calculated $\alpha$ decay half-lives and the effect of the laser field with an intensity of $10^{23}\,\mathrm{W}/\mathrm{cm}^{2}$ on the ground-state $\alpha$ decay of odd-odd nuclei.}
\label{tab1}
\begin{tabular*}{\textwidth}{@{\extracolsep{\fill}}lcccccc@{\quad}lcccccc@{}}
\CPCtoprule
Nucleus & $Q_\alpha$ (MeV) & $l$ & ${\log_{10}} T_{\rm exp}$ (s) & ${\log_{10}} T_{\rm cal}$ (s) & $\Delta P^{23}$ & $\Delta T^{23}$ &
Nucleus & $Q_\alpha$ (MeV) & $l$ & ${\log_{10}} T_{\rm exp}$ (s) & ${\log_{10}} T_{\rm cal}$ (s) & $\Delta P^{23}$ & $\Delta T^{23}$\\
\CPCmidrule
$^{110}$I&3.580&2&0.582&{ 0.559}&{ $1.756\times 10^{-4}$}&{ $-1.756\times 10^{-4}$} & $^{210}$At&5.631&2&7.221&{ 6.787}&{ $2.863\times 10^{-4}$}&{ $-2.862\times 10^{-4}$}\\
$^{112}$I&2.957&4&5.445&{ 5.942}&{ $2.715\times 10^{-4}$}&{ $-2.714\times 10^{-4}$} & $^{212}$At&7.817&5&-0.503&{ -1.246}&{ $1.548\times 10^{-4}$}&{ $-1.548\times 10^{-4}$}\\
$^{114}$Cs&3.360&0&3.486&{ 2.972}&{ $2.111\times 10^{-4}$}&{ $-2.110\times 10^{-4}$} & $^{214}$At&8.988&0&-6.253&{ -5.918}&{ $1.144\times 10^{-4}$}&{ $-1.143\times 10^{-4}$}\\
$^{152}$Ho&4.507&0&3.119&{ 2.816}&{ $2.378\times 10^{-4}$}&{ $-2.378\times 10^{-4}$} & $^{216}$At&7.950&0&-3.523&{ -3.083}&{ $1.542\times 10^{-4}$}&{ $-1.542\times 10^{-4}$}\\
$^{154}$Ho&4.041&0&6.559&{ 5.649}&{ $3.020\times 10^{-4}$}&{ $-3.019\times 10^{-4}$} & $^{218}$At&6.876&0&0.176&{ 0.546}&{ $2.049\times 10^{-4}$}&{ $-2.049\times 10^{-4}$}\\
$^{154}$Tm&5.094&0&1.172&{ 0.900}&{ $1.862\times 10^{-4}$}&{ $-1.861\times 10^{-4}$} & $^{200}$Fr&7.622&0&-1.323&{ -0.720}&{ $1.401\times 10^{-4}$}&{ $-1.401\times 10^{-4}$}\\
$^{156}$Tm&4.345&0&5.116&{ 4.945}&{ $2.662\times 10^{-4}$}&{ $-2.661\times 10^{-4}$} & $^{204}$Fr&7.170&0&0.260&{ 0.831}&{ $1.635\times 10^{-4}$}&{ $-1.635\times 10^{-4}$}\\
$^{158}$Lu&4.790&0&3.065&{ 3.549}&{ $2.173\times 10^{-4}$}&{ $-2.172\times 10^{-4}$} & $^{206}$Fr&6.923&0&1.258&{ 1.748}&{ $1.833\times 10^{-4}$}&{ $-1.833\times 10^{-4}$}\\
$^{158}$Ta&6.124&0&-1.294&{ -1.483}&{ $1.327\times 10^{-4}$}&{ $-1.327\times 10^{-4}$} & $^{208}$Fr&6.785&0&1.821&{ 2.271}&{ $1.909\times 10^{-4}$}&{ $-1.909\times 10^{-4}$}\\
$^{162}$Ta&5.010&5&3.683&{ 5.091}&{ $2.089\times 10^{-4}$}&{ $-2.088\times 10^{-4}$} & $^{210}$Fr&6.671&2&2.427&{ 3.009}&{ $2.014\times 10^{-4}$}&{ $-2.013\times 10^{-4}$}\\
$^{160}$Re&6.698&2&-2.259&{ -2.325}&{ $1.145\times 10^{-4}$}&{ $-1.145\times 10^{-4}$} & $^{212}$Fr&6.529&2&3.444&{ 3.583}&{ $2.115\times 10^{-4}$}&{ $-2.115\times 10^{-4}$}\\
$^{162}$Re&6.240&0&-0.947&{ -0.989}&{ $1.330\times 10^{-4}$}&{ $-1.330\times 10^{-4}$} & $^{214}$Fr&8.589&5&-2.286&{ -2.728}&{ $1.308\times 10^{-4}$}&{ $-1.308\times 10^{-4}$}\\
$^{166}$Re&5.520&0&1.636&{ 2.026}&{ $1.832\times 10^{-4}$}&{ $-1.832\times 10^{-4}$} & $^{216}$Fr&9.174&0&-6.155&{ -5.637}&{ $1.142\times 10^{-4}$}&{ $-1.142\times 10^{-4}$}\\
$^{168}$Re&5.063&4&4.944&{ 5.340}&{ $2.209\times 10^{-4}$}&{ $-2.209\times 10^{-4}$} & $^{218}$Fr&8.014&0&-3.000&{ -2.457}&{ $1.506\times 10^{-4}$}&{ $-1.506\times 10^{-4}$}\\
$^{166}$Ir&6.722&0&-1.949&{ -1.851}&{ $1.195\times 10^{-4}$}&{ $-1.195\times 10^{-4}$} & $^{220}$Fr&6.801&1&1.438&{ 1.847}&{ $2.108\times 10^{-4}$}&{ $-2.108\times 10^{-4}$}\\
$^{168}$Ir&6.381&0&-0.638&{ -0.607}&{ $1.389\times 10^{-4}$}&{ $-1.389\times 10^{-4}$} & $^{206}$Ac&7.960&0&-1.602&{ -1.009}&{ $1.386\times 10^{-4}$}&{ $-1.386\times 10^{-4}$}\\
$^{172}$Ir&5.991&3&2.342&{ 1.541}&{ $1.675\times 10^{-4}$}&{ $-1.675\times 10^{-4}$} & $^{216}$Ac&9.241&5&-3.357&{ -3.714}&{ $1.149\times 10^{-4}$}&{ $-1.148\times 10^{-4}$}\\
$^{174}$Ir&5.693&2&3.173&{ 2.527}&{ $1.898\times 10^{-4}$}&{ $-1.897\times 10^{-4}$} & $^{218}$Ac&9.380&0&-6.000&{ -5.420}&{ $1.114\times 10^{-4}$}&{ $-1.114\times 10^{-4}$}\\
$^{170}$Au&7.177&0&-2.583&{ -2.522}&{ $1.126\times 10^{-4}$}&{ $-1.126\times 10^{-4}$} & $^{220}$Ac&8.348&2&-1.579&{ -2.380}&{ $1.441\times 10^{-4}$}&{ $-1.441\times 10^{-4}$}\\
$^{176}$Au&6.433&0&0.142&{ 0.082}&{ $1.516\times 10^{-4}$}&{ $-1.516\times 10^{-4}$} & $^{222}$Ac&7.137&0&0.703&{ 1.367}&{ $1.970\times 10^{-4}$}&{ $-1.970\times 10^{-4}$}\\
$^{180}$Tl&6.710&0&1.231&{ -0.053}&{ $1.461\times 10^{-4}$}&{ $-1.461\times 10^{-4}$} & $^{224}$Ac&6.327&1&5.026&{ 4.792}&{ $2.534\times 10^{-4}$}&{ $-2.533\times 10^{-4}$}\\
$^{186}$Bi&7.757&1&-1.830&{ -2.662}&{ $1.180\times 10^{-4}$}&{ $-1.180\times 10^{-4}$} & $^{226}$Ac&5.506&2&9.232&{ 9.120}&{ $3.394\times 10^{-4}$}&{ $-3.393\times 10^{-4}$}\\
$^{190}$Bi&6.862&1&0.912&{ 0.327}&{ $1.581\times 10^{-4}$}&{ $-1.581\times 10^{-4}$} & $^{212}$Pa&8.410&4&-2.125&{ -0.676}&{ $1.282\times 10^{-4}$}&{ $-1.282\times 10^{-4}$}\\
$^{192}$Bi&6.377&1&2.442&{ 2.217}&{ $1.842\times 10^{-4}$}&{ $-1.841\times 10^{-4}$} & $^{220}$Pa&9.704&0&-6.108&{ -5.479}&{ $1.063\times 10^{-4}$}&{ $-1.063\times 10^{-4}$}\\
$^{194}$Bi&5.918&1&4.313&{ 4.230}&{ $2.143\times 10^{-4}$}&{ $-2.143\times 10^{-4}$} & $^{224}$Pa&7.694&2&-0.073&{ 0.508}&{ $1.713\times 10^{-4}$}&{ $-1.712\times 10^{-4}$}\\
$^{212}$Bi&6.207&5&4.005&{ 3.799}&{ $2.380\times 10^{-4}$}&{ $-2.379\times 10^{-4}$} & $^{228}$Pa&6.264&3&6.631&{ 6.520}&{ $2.675\times 10^{-4}$}&{ $-2.674\times 10^{-4}$}\\
$^{214}$Bi&5.621&5&6.753&{ 6.536}&{ $3.440\times 10^{-4}$}&{ $-3.439\times 10^{-4}$} & $^{230}$Pa&5.439&2&10.671&{ 10.566}&{ $3.569\times 10^{-4}$}&{ $-3.568\times 10^{-4}$}\\
$^{192}$At&7.696&0&-1.939&{ -1.768}&{ $1.277\times 10^{-4}$}&{ $-1.277\times 10^{-4}$} & $^{236}$Np&5.010&4&15.469&{ 15.043}&{ $4.516\times 10^{-4}$}&{ $-4.514\times 10^{-4}$}\\
$^{200}$At&6.596&0&1.917&{ 2.132}&{ $1.849\times 10^{-4}$}&{ $-1.849\times 10^{-4}$} & $^{236}$Am&6.260&1&6.832&{ 8.040}&{ $2.805\times 10^{-4}$}&{ $-2.804\times 10^{-4}$}\\
$^{202}$At&6.354&0&3.161&{ 3.131}&{ $2.054\times 10^{-4}$}&{ $-2.053\times 10^{-4}$} & $^{244}$Bk&6.779&2&8.458&{ 6.717}&{ $2.536\times 10^{-4}$}&{ $-2.536\times 10^{-4}$}\\
$^{204}$At&6.070&0&4.156&{ 4.381}&{ $2.290\times 10^{-4}$}&{ $-2.290\times 10^{-4}$} & $^{248}$Es&7.160&2&5.760&{ 5.977}&{ $2.358\times 10^{-4}$}&{ $-2.358\times 10^{-4}$}\\
$^{206}$At&5.887&0&5.306&{ 5.234}&{ $2.484\times 10^{-4}$}&{ $-2.483\times 10^{-4}$} & $^{258}$Md&7.271&1&6.648&{ 6.181}&{ $2.458\times 10^{-4}$}&{ $-2.458\times 10^{-4}$}\\
$^{208}$At&5.751&0&6.023&{ 5.888}&{ $2.616\times 10^{-4}$}&{ $-2.616\times 10^{-4}$} & $^{272}$Rg&11.197&0&-2.347&{ -2.328}&{ $1.136\times 10^{-4}$}&{ $-1.136\times 10^{-4}$}\\
\CPCbottomrule
\end{tabular*}
\end{table*}

\subsubsection{Laser-nucleus interaction}

The laser electric field is represented by a linearly polarized Gaussian plane wave, written as
\begin{equation}
E(t)=E_0 f(t) \sin(\omega t), 
\end{equation}
where $\omega$ represents the angular frequency. The peak laser electric field is related to the peak intensity $I_0$ by $E_0\,[\mathrm{V\,cm^{-1}}]=27.44\left(I_0\,[\,{\mathrm{W}/\mathrm{cm}^{2}}]\right)^{1/2}$ \cite{Mi_icu_2019}. $f(t)$ is the Gaussian pulse envelope, written as
\begin{equation}
f(t)=\exp\left(-\frac{t^2}{\tau^2}\right).
\end{equation}
The pulse width of the envelope $\tau$ is expressed in terms of the pulse period $T_0$ as
\begin{equation}
\tau=x T_0.
\end{equation}
Combined with the expression for the electric field, $E(t)$ can be written in a wavelength-dependent form
\begin{equation}
\label{eq:laser-field}
\begin{aligned}
E(t)&=E_0\exp\left(-\frac{t^2}{x^2T_0^2}\right)\sin(\omega t)\\
&=E_0\exp\left(-\frac{c^2t^2}{\lambda^2x^2}\right)\sin\left(2\pi\frac{c}{\lambda}t\right).
\end{aligned}
\end{equation}
where $\lambda$ represents the laser wavelength. The pulse period is $T_0=1/\nu$, where $\nu=\omega/2\pi=c/\lambda$ is the laser frequency. 

In the interaction between laser electric fields and nuclear decay, the influence of the electric field on the decay energy $Q_{\alpha}$ should be considered. The change in the decay energy $\Delta Q_{\alpha}$ is equivalent to the energy gained by the $\alpha$ particle from the laser electric field while it penetrates the potential barrier. It is expressed as
\begin{equation}
\Delta Q_{\alpha}=eZ_\alpha E(t)R_d(\varphi)\cos\theta.
\end{equation}
The $\alpha$ decay energy including the laser electric-field effect, $Q^*_{\alpha}$, can thus be rewritten as
\begin{equation}
Q^*_{\alpha}=Q_{\alpha}+\Delta Q_{\alpha}.
\label{eq28}
\end{equation}
To quantify the influence of the laser-modified decay energy and potential barrier on the decay process, the relative change in the penetration probability is defined as
\begin{equation}
\Delta P=
\frac{P(E,\theta)-P(E=0,\theta)}
{P(E=0,\theta)},
\label{eq29}
\end{equation}
where $P(E,\theta)$ and $P(E=0,\theta)$ denote the penetration probabilities with and without the laser electric field, respectively. Correspondingly, the relative change in the $\alpha$ decay half-life is written as
\begin{equation}
\Delta T=
\frac{T(E,\theta)-T(E=0,\theta)}
{T(E=0,\theta)} .
\label{eq30}
\end{equation}
Since the $\alpha$ decay half-life is inversely proportional to the penetration probability when the other factors in the decay width are fixed, $\Delta T$ can be related to $\Delta P$ as
\begin{equation}
\Delta T=
\frac{P(E=0,\theta)-P(E,\theta)}
{P(E,\theta)}
=
-\frac{\Delta P}{1+\Delta P}.
\label{eq31}
\end{equation}
Therefore, a positive $\Delta P$ corresponds to an enhanced barrier penetration probability and a shortened $\alpha$ decay half-life. In the present work, $\Delta P$ is used as the main quantity to characterize the laser-induced modification, while $\Delta T$ is given to show the corresponding change in the half-life.

\section{Results and discussion}\label{Sec.III}

\subsection{Accuracy of the calculated \texorpdfstring{$\alpha$}{alpha} decay half-lives}

Before analyzing the laser-induced change of odd-odd $\alpha$ decay, we first examine whether the adopted deformed cluster model gives a reliable description of the known half-lives without the external laser field. This step is essential because the quantities $\Delta P$ and $\Delta T$ are obtained by comparing the barrier penetration probability with and without the laser field. Therefore, an inaccurate zero-field half-life would directly weaken the physical meaning of the calculated laser response. Following the same logic used in our previous calculations for even-even and odd-$A$ nuclei, the potential parameters are determined by reproducing the experimental ground-state $\alpha$ decay half-lives of the selected odd-odd nuclei \cite{CHENG2024138322, cjh2025}.

{The deformation parameters required for the calculations can be obtained from various theoretical studies, such as DRHBc studies of nuclear stability \cite{Zhang2021DRHBcPeninsula,He2024DRHBcPeninsula} and recent mass-table developments \cite{Zhang2025DRHBcReview,Guo2026DRHBcIII}, as well as the microscopic--macroscopic FRDM2012 model \cite{MOLLER20161}.} {Since the present calculation requires the} deformation parameters $\beta_2$, $\beta_4$, and $\beta_6${, we take these parameters consistently from FRDM2012 to maintain uniform deformation inputs.} The $\alpha$ decay energies, spin-parity assignments, and experimental half-lives are taken from AME2020 and NUBASE2020 \cite{CPC-2021-0034, CPC-2020-0033, NUBASE2020}.

The agreement between the calculated and experimental half-lives is quantified by the logarithmic root-mean-square deviation
\begin{equation}
\Delta=\sqrt{\frac{1}{n}\sum_{i=1}^{n}\left[\log_{10}T_{1/2,i}^{\rm exp}(s)-\log_{10}T_{1/2,i}^{\rm cal}(s)\right]^2}.
\label{eq32}
\end{equation}
For the present odd-odd-nucleus set, we obtain $\Delta=0.571$. Because $\alpha$ decay half-lives span many orders of magnitude, this logarithmic deviation is a direct and meaningful measure of the global predictive accuracy of the model.

The calculated results are listed in Table \ref{tab1}. The table includes the parent nucleus, $\alpha$ decay energy $Q_\alpha$, orbital angular momentum $l$ carried by the emitted $\alpha$ particle, experimental and calculated logarithmic half-lives, and the laser-induced relative changes $\Delta P^{23}$ and $\Delta T^{23}$. Here, the superscript 23 denotes the peak laser intensity of $10^{23}\,\mathrm{W}/\mathrm{cm}^{2}$. As shown in Table \ref{tab1}, the calculated half-lives reproduce the experimental values over a wide range from short-lived neutron-deficient nuclei to long-lived actinide-region nuclei. This indicates that the dominant effects of $Q_\alpha$, deformation, and angular-momentum transfer are reasonably included in the present deformed barrier calculation.

\begin{figure}[!tbp]
\centering
\includegraphics[width=0.95\columnwidth]{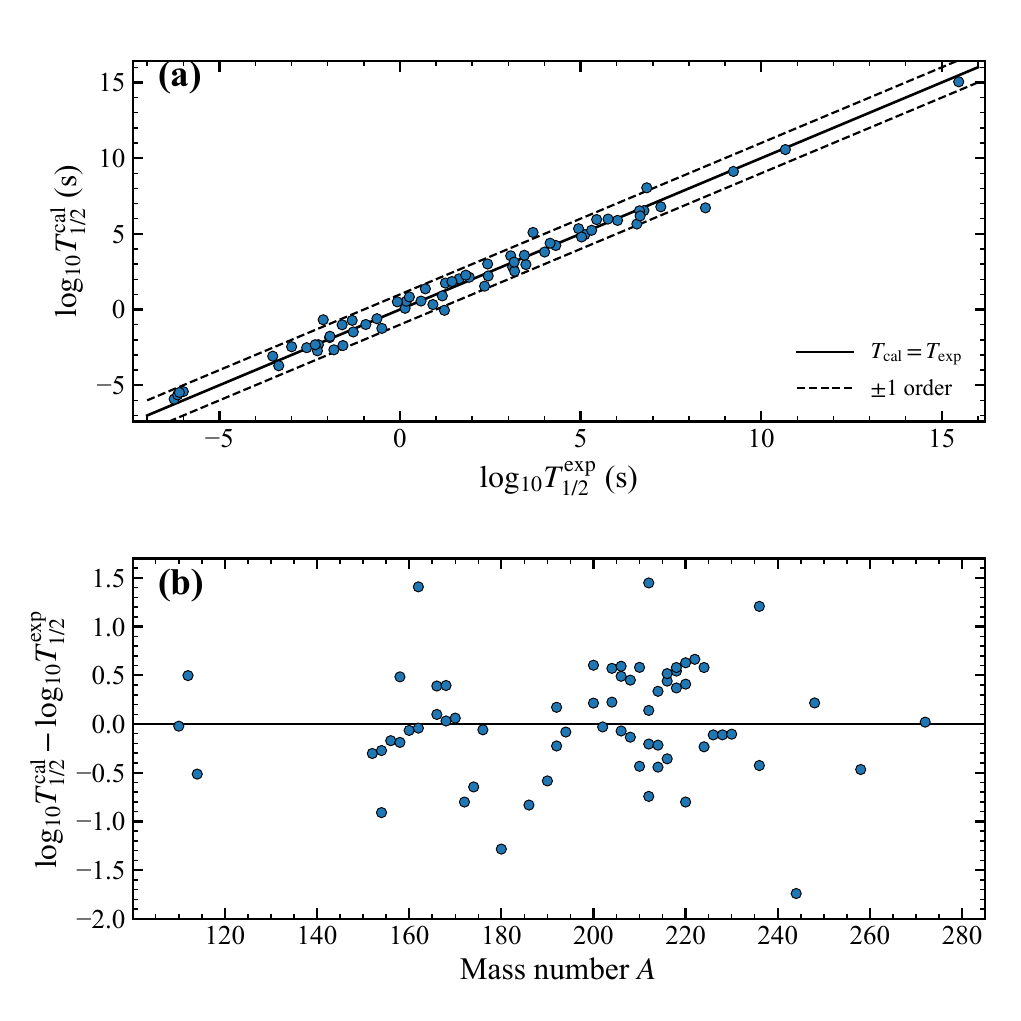}
\caption{Comparison between the calculated and experimental $\alpha$ decay half-lives of the selected odd-odd nuclei. (a) Correlation between $\log_{10}T_{1/2}^{\rm exp}$ and $\log_{10}T_{1/2}^{\rm cal}$. The solid line represents $T_{\rm cal}=T_{\rm exp}$, and the dashed lines denote deviations of one order of magnitude. (b) Residuals between the calculated and experimental logarithmic half-lives as a function of the mass number $A$.}
\label{fig2}
\end{figure}

The comparison between the calculated and experimental half-lives is further illustrated in Fig. \ref{fig2}. In Fig. \ref{fig2}(a), the calculated values are plotted as a function of the experimental values. The solid diagonal line represents $T_{\rm cal}=T_{\rm exp}$, while the two dashed lines correspond to deviations of one order of magnitude. Most of the points lie close to the diagonal line, which demonstrates that the fitted potential can follow the main half-life trend of odd-odd nuclei. Fig. \ref{fig2}(b) shows the residual $\log_{10}T_{1/2}^{\rm cal}-\log_{10}T_{1/2}^{\rm exp}$ as a function of the mass number $A$. The residuals fluctuate around zero and no obvious monotonic drift appears with increasing $A$. This behavior shows that the present parameter set does not simply fit one local mass region, but gives a reasonably balanced description for the selected odd-odd nuclei from the medium-heavy region to the actinide region.

Overall, the calculated $\alpha$ decay half-lives are in good agreement with the experimental ones. The value of $\Delta$ is comparable to the accuracy usually obtained in systematic calculations of unfavored and structure-dependent $\alpha$ decay, especially for odd-odd nuclei where the unpaired proton and unpaired neutron introduce additional hindrance and configuration dependence. Therefore, the present deformed density-dependent cluster model provides a reliable baseline for the following discussion of laser-induced changes in the penetration probability and half-life.

\subsection{Laser-induced variation and the branch structure of odd-odd nuclei}

After establishing the reliability of the zero-field half-life calculation, we now discuss the influence of the ultra-intense laser field on the $\alpha$ decay penetration probability. In the present calculations, the peak intensity is fixed at $10^{23}\,\mathrm{W}/\mathrm{cm}^{2}$ ($\cos\theta=1$), and the relative changes of the penetration probability and half-life are denoted by $\Delta P^{23}$ and $\Delta T^{23}$, respectively. This fixed-intensity treatment is useful because it removes the ambiguity caused by different laser parameters and allows the nuclear-structure dependence to be examined directly. Since the decay width is proportional to the penetration probability in the present framework, the half-life is inversely related to $P$. Therefore, the two relative changes are connected through
\begin{equation}
\Delta T^{23}=-\frac{\Delta P^{23}}{1+\Delta P^{23}}.
\label{eq33}
\end{equation}
All calculated $\Delta P^{23}$ values are much smaller than unity. The maximum value of $|\Delta T^{23}+\Delta P^{23}|/\Delta P^{23}$ is only ${4.5\times10^{-4}}$ for the 66 odd-odd nuclei, which means that $\Delta T^{23}\simeq-\Delta P^{23}$ is sufficiently accurate for the present discussion. In other words, a nucleus with a larger positive $\Delta P^{23}$ corresponds directly to a larger relative shortening of the $\alpha$ decay half-life. This simple relation is important for experiments, because the penetration-probability variation can be used as an equivalent indicator for ranking candidate nuclei according to their laser sensitivity.

The first systematic feature is the negative correlation between $Q_{\alpha}$ and $\Delta P^{23}$. To make the statistical statement clear, we use the Pearson correlation coefficient $r$ to quantify the linear correlation between two variables. The coefficient $r$ is the covariance of the two variables divided by the product of their standard deviations. Thus, $r=-1$ represents a perfect negative linear relation, $r=1$ a perfect positive linear relation, and $r=0$ no linear relation. For all selected odd-odd nuclei, we obtain $r(Q_{\alpha},\Delta P^{23})=-0.620$ and $r(Q_{\alpha},\log_{10}\Delta P^{23})=-0.680$. The negative sign means that nuclei with smaller $Q_{\alpha}$ values tend to be more sensitive to the external laser field. This behavior is consistent with the tunneling picture obtained in even-even and odd-$A$ nuclei: when the decay energy is lower, the emitted $\alpha$ particle samples a wider and more sensitive barrier region, and a small laser-induced modification of the effective potential is amplified in the relative penetration probability.

However, the odd-odd systematics cannot be reduced to the $Q_{\alpha}$ effect alone. A single-variable fit of $\log_{10}\Delta P^{23}$ as a function of $Q_{\alpha}$ gives only $R^2=0.462$. Here $R^2$, the coefficient of determination, measures how much of the data variance is explained by a given fitting model. A value close to unity means that the model explains most of the systematic trend, whereas a small value indicates that important variables are missing. The moderate value obtained from the $Q_{\alpha}$-only fit therefore tells us that decay energy is the leading energy-scale parameter, but additional odd-odd structural information is required to understand the dispersion of the laser response.

\begin{figure}[!tbp]
\centering
\includegraphics[width=0.98\columnwidth]{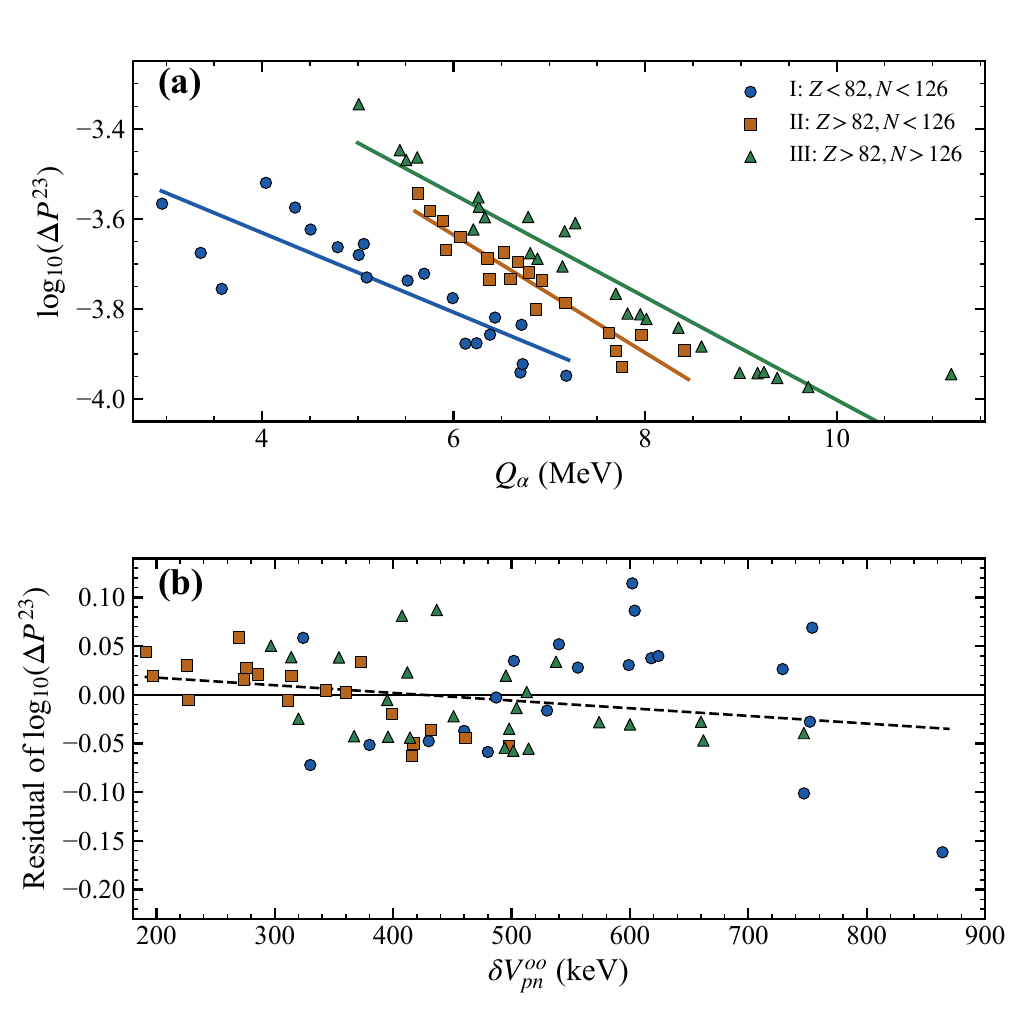}
\caption{Systematics of the laser-induced penetration-probability variation in odd-odd nuclei. (a) $\log_{10}(\Delta P^{23})$ as a function of $Q_{\alpha}$, where the nuclei are divided into three branches according to their positions relative to $Z=82$ and $N=126$. The solid lines are independent linear fits for the three branches. (b) Residual of $\log_{10}(\Delta P^{23})$ after removing the $Q_{\alpha}$ dependence and the three-branch classification, plotted as a function of the empirical residual proton-neutron interaction $\delta V_{pn}^{oo}$.}
\label{fig3}
\end{figure}

A remarkable feature emerges when the odd-odd nuclei are classified according to their positions relative to the major shell closures $Z=82$ and $N=126$. As shown in Fig. \ref{fig3}(a) and Table \ref{tab2}, the $\log_{10}\Delta P^{23}$--$Q_{\alpha}$ systematics splits into three clear branches: $Z<82,N<126$, $Z>82,N<126$, and $Z>82,N>126$. This classification reflects whether the unpaired proton and the unpaired neutron behave as particles or holes with respect to the shell closures. After this classification, the fitted $R^2$ values increase to {0.721, 0.898, and 0.902} for the three branches, respectively. This strong improvement from the global $R^2=0.462$ demonstrates that the simultaneous presence of an unpaired proton and an unpaired neutron leaves a visible imprint on the laser-induced tunneling response. The branch structure is therefore the main additional physics of the present odd-odd study. It is different from the analytical $Q_{\alpha}$-dependent estimate emphasized in even-even nuclei and also different from the odd-even staggering discussed in odd-$A$ nuclei, because it originates from the coupled valence proton-neutron particle-hole configuration that only appears in odd-odd systems.

\begin{table*}[!t]
\centering
\footnotesize
\setlength{\tabcolsep}{3.0pt}
\renewcommand{\arraystretch}{1.12}
\caption{Branch-dependent systematics of $\log_{10}(\Delta P^{23})$ as a function of $Q_{\alpha}$ in odd-odd nuclei. The fitted relation is $\log_{10}(\Delta P^{23})=aQ_{\alpha}+b$.}
\label{tab2}
\begin{tabular*}{\textwidth}{@{\extracolsep{\fill}}llcccccc@{}}
\CPCtoprule
Branch & Valence type & $n$ & $Q_{\alpha}$ range (MeV) & $\Delta P^{23}$ range & $r$ & Fitted parameters & $R^2$\\
\CPCmidrule
I: $Z<82,N<126$&hole-hole&21&2.957--7.177&{$(1.126$--$3.020)\times10^{-4}$}&{-0.849}&\makecell{{$a=-0.088$}\\{$b=-3.278$}}&{0.721}\\
II: $Z>82,N<126$&particle-hole&19&5.631--8.410&{$(1.180$--$2.863)\times10^{-4}$}&{-0.948}&\makecell{{$a=-0.131$}\\{$b=-2.851$}}&{0.898}\\
III: $Z>82,N>126$&particle-particle&26&5.010--11.197&{$1.063\times10^{-4}$--$4.516\times10^{-4}$}&{-0.950}&\makecell{{$a=-0.114$}\\{$b=-2.861$}}&{0.902}\\
\CPCbottomrule
\end{tabular*}
\end{table*}

\begin{figure*}[!tbp]
\centering
\includegraphics[width=0.8\textwidth]{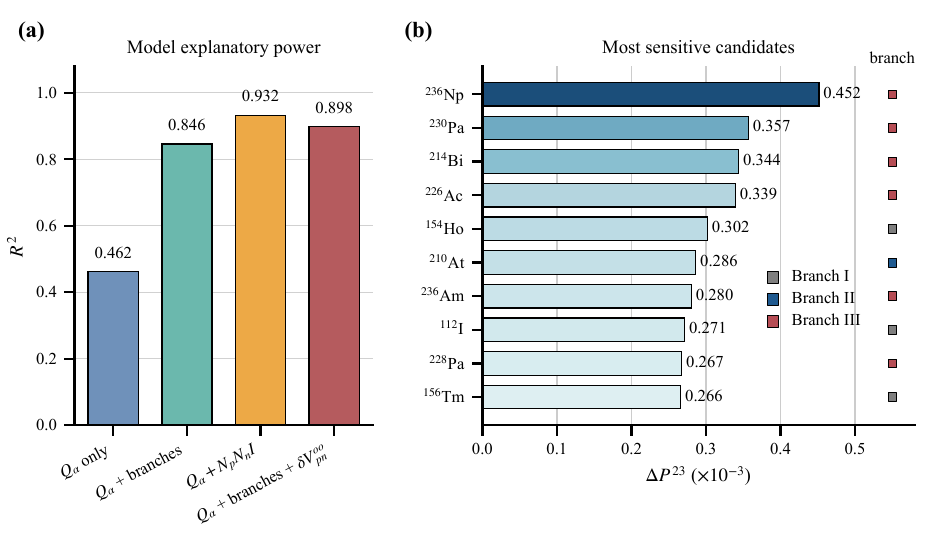}
\caption{Role of odd-odd structural variables and candidate nuclei for future laser-assisted $\alpha$ decay studies. (a) Comparison of the explanatory power of different descriptions for $\log_{10}(\Delta P^{23})$. The large increase in $R^2$ after including the branch structure and the valence quantities demonstrates the importance of the unpaired proton-neutron configuration. (b) Ten nuclei with the largest $\Delta P^{23}$ values at $10^{23}\,\mathrm{W}/\mathrm{cm}^{2}$. The colors denote the three branch regions.}
\label{fig4}
\end{figure*}

The origin of the branch structure can be further understood from the laser-nucleus coupling term and the valence proton-neutron configuration. For $\alpha$ decay, the effective charge in the relative coordinate reduces to
\begin{equation}
\begin{aligned}
\frac{Q_{\rm eff}}{e}
&=\frac{Z_{\alpha}A_d-Z_dA_{\alpha}}{A}\\
&=\frac{2(A-4)-4(Z-2)}{A}\\
&=\frac{2(N-Z)}{A}=2I.
\end{aligned}
\label{eq34}
\end{equation}
where $I=(N-Z)/A$ is the isospin asymmetry. This relation shows that $I$ is not a purely empirical fitting quantity; it enters directly into the strength with which the laser electric field separates the $\alpha$ particle and the daughter nucleus in the relative coordinate. We therefore introduce the valence quantities $N_p=|Z-82|$ and $N_n=|N-126|$ to characterize the positions of the unpaired proton and neutron relative to the shell closures. When $Q_{\alpha}$, $N_p$, $N_n$, and $I$ are used together to describe $\log_{10}\Delta P^{23}$, the coefficient of determination increases to $R^2={0.932}$, as illustrated in Fig. \ref{fig4}(a). This result supports the view that the laser sensitivity of odd-odd nuclei is governed by both the tunneling energy scale and the unpaired proton-neutron structure.

To obtain an independent indicator of the residual proton-neutron correlation in odd-odd nuclei, we also calculate the empirical interaction energy from AME2020 mass differences. For an odd-odd nucleus with proton number $Z$ and neutron number $N$, it is defined by the binding-energy double difference \cite{Wu2016EmpiricalPn,Fu2013RegularitiesPN}
\begin{equation}
\begin{aligned}
\delta V_{pn}^{oo}(Z,N)={}&B(Z,N)-B(Z,N-1)\\
&-B(Z-1,N)+B(Z-1,N-1).
\end{aligned}
\label{eq35}
\end{equation}
Equivalently, using mass excess values from AME2020, the constants associated with proton and neutron masses cancel and Eq. (\ref{eq35}) can be rewritten as
\begin{equation}
\begin{aligned}
\delta V_{pn}^{oo}={}&-ME(Z,N)+ME(Z,N-1)\\
&+ME(Z-1,N)-ME(Z-1,N-1).
\end{aligned}
\label{eq36}
\end{equation}
This quantity measures the additional average proton-neutron interaction associated with the simultaneous odd proton and odd neutron. It should not be regarded as the first-order determinant of $\Delta P^{23}$, because $Q_{\alpha}$ and the shell branch dominate the systematics. Indeed, the direct correlation between $\delta V_{pn}^{oo}$ and $\log_{10}\Delta P^{23}$ is weak. Nevertheless, after the $Q_{\alpha}$ dependence and the three-branch structure are subtracted, the residual shows a moderate correlation with $\delta V_{pn}^{oo}$, as shown in Fig. \ref{fig3}(b). In this context, the residual means the difference between the calculated value of $\log_{10}\Delta P^{23}$ and the value predicted by the fitted $Q_{\alpha}$ plus branch model. It is useful because it highlights the part of the laser response that remains unexplained after the dominant energy and shell-classification effects have been removed. For the 65 nuclei for which $\delta V_{pn}^{oo}$ can be obtained, adding this quantity improves $R^2$ from {0.868 to 0.898}. Therefore, $\delta V_{pn}^{oo}$ provides a useful secondary validation that the remaining dispersion contains information on the unpaired proton-neutron correlation, rather than only numerical noise.

The above analysis clarifies why the odd-odd case is not merely a larger data set following the even-even or odd-$A$ trend. The effective charge, the valence distances $N_p$ and $N_n$, and the empirical residual interaction probe different aspects of the same physical problem: how the unpaired proton and neutron modify the decay barrier and the response of the relative $\alpha$-daughter coordinate to the laser electric field. The improvement of the statistical description after adding these variables should be understood as evidence for a structure-dependent laser response, not as a purely mathematical fit. The branch structure is useful not only for understanding the systematics, but also for selecting candidate nuclei in future laser-assisted decay experiments. Table \ref{tab3} lists the ten nuclei with the largest calculated $\Delta P^{23}$. The largest value appears in $^{236}$Np, followed by $^{230}$Pa and {$^{214}$Bi}. These nuclei are mainly located in the $Z>82,N>126$ branch, where the combined influence of relatively low $Q_{\alpha}$ values and valence proton-neutron structure leads to an enhanced laser sensitivity. The appearance of $^{154}$Ho and $^{112}$I in this list also indicates that a low decay energy can compensate for belonging to a different shell branch, which is consistent with the conclusion that $Q_{\alpha}$ is the leading but not exclusive control parameter.

From an experimental perspective, the present ranking provides a first theoretical filter for choosing candidate sources. A larger $\Delta P^{23}$ means a larger expected relative half-life shortening at the same peak laser intensity, and therefore a more favorable signal size. At the same time, practical feasibility must still be judged by production yield, source preparation, $\alpha$ decay branch ratio, background control, detector geometry, and the half-life window accessible in repeated laser-shot measurements. The present analysis narrows the nuclear candidates before these experimental constraints are applied: it identifies which odd-odd nuclei are intrinsically more sensitive to the laser field and explains why this sensitivity is enhanced. In this sense, the three-branch systematics and the valence proton-neutron indicators provide not only a structural interpretation but also a practical guide for designing future laser-assisted $\alpha$ decay experiments.

\begin{table}[!tbp]
\centering
\scriptsize
\setlength{\tabcolsep}{0.5pt}
\renewcommand{\arraystretch}{1.35}
\caption{Candidate nuclei with the largest laser-induced relative change of the penetration probability at $10^{23}\,\mathrm{W}/\mathrm{cm}^{2}$. Here, Br. denotes branch.}
\label{tab3}
\begin{tabular*}{\columnwidth}{@{\extracolsep{\fill}}lcccccc@{}}
\CPCtoprule
Nucleus & $Q_{\alpha}$ (MeV) & $\Delta P^{23}$ & $\Delta T^{23}$ & Br. & $N_pN_nI$ & $\delta V_{pn}^{oo}$ (keV)\\
\CPCmidrule
${}^{236}$Np&5.010&{ $4.516\times10^{-4}$}&{ $-4.514\times10^{-4}$}&III&39.62&436.8\\
${}^{230}$Pa&5.439&{ $3.569\times10^{-4}$}&{ $-3.568\times10^{-4}$}&III&24.42&537.6\\
${}^{214}$Bi&5.621&{ $3.440\times10^{-4}$}&{ $-3.439\times10^{-4}$}&III&1.12&313.9\\
${}^{226}$Ac&5.506&{ $3.394\times10^{-4}$}&{ $-3.393\times10^{-4}$}&III&16.35&495.2\\
${}^{154}$Ho&4.041&{ $3.020\times10^{-4}$}&{ $-3.019\times10^{-4}$}&I&75.97&602.0\\
${}^{210}$At&5.631&{ $2.863\times10^{-4}$}&{ $-2.862\times10^{-4}$}&II&0.57&191.2\\
${}^{236}$Am&6.260&{ $2.805\times10^{-4}$}&{ $-2.804\times10^{-4}$}&III&38.01&412.0\\
${}^{112}$I&2.957&{ $2.715\times10^{-4}$}&{ $-2.714\times10^{-4}$}&I&104.09&752.0\\
${}^{228}$Pa&6.264&{ $2.675\times10^{-4}$}&{ $-2.674\times10^{-4}$}&III&19.97&512.8\\
{${}^{156}$Tm}&4.345&{ $2.662\times10^{-4}$}&{ $-2.661\times10^{-4}$}&I&58.50&604.0\\
\CPCbottomrule
\end{tabular*}
\end{table}

\subsection{Enhancement of the pulse-averaged laser response using two-color pulses}

The above discussion identifies the odd-odd nuclei that are intrinsically sensitive to an ultra-intense laser field. However, the quantities listed in Table \ref{tab1} and used in the branch analysis describe the response under a fixed field condition, whereas a real experiment is performed with a finite laser pulse. Therefore, the complete-pulse effect is more relevant for designing a measurable laser-assisted decay experiment. The laser pulse has a finite duration and a specific temporal profile, so the electric field affects $\Delta P$ differently at different moments. Since the laser electric field oscillates in time, the field-odd part of the penetration-probability variation is largely cancelled in one complete pulse. This means that the pulse-averaged relative change in the $\alpha$ decay penetration probability can be much smaller than the maximum instantaneous relative change.

This cancellation can be understood from the analytical expression obtained in our previous work \cite{CHENG2024138322}. In the Gamow-like approximation, the relative change in the penetration probability can be separated into first-order and second-order laser-field contributions,
\begin{equation}
\Delta P=\chi^{(1)}+\chi^{(2)}.
\label{eq37}
\end{equation}
The first-order term can be approximated as
\begin{equation}
\begin{aligned}
\chi^{(1)} \approx{}& B_1 Q_\alpha^{-5/2}
+B_2 Q_\alpha^{-2}+B_3 Q_\alpha^{-1},\\
B_1={}&\frac{3\pi E(t)\sqrt{2\mu}Q_{\rm eff}\cos\theta}{8\hbar}
(Z_dZ_\alpha e^2)^2,\\
B_2={}&\frac{3E(t)\sqrt{2\mu}Q_{\rm eff}\cos\theta}{4\hbar}
(Z_dZ_\alpha e^2)^{3/2}R_{\rm in}^{1/2},\\
B_3={}&\frac{E(t)\sqrt{2\mu}Q_{\rm eff}\cos\theta}{2\hbar}
(Z_dZ_\alpha e^2)^{1/2}R_{\rm in}^{3/2}.
\end{aligned}
\label{eq37a}
\end{equation}
The second-order term is
\begin{equation}
\begin{aligned}
\chi^{(2)} \approx{}& C_1 Q_\alpha^{-9/2},\\
C_1={}&\frac{E^2(t)\sqrt{2\mu}(Q_{\rm eff}\cos\theta)^2}{4\hbar}
(Z_dZ_\alpha e^2)^3
\left(T_{rc}-\frac{35\pi}{16}\right).
\end{aligned}
\label{eq37b}
\end{equation}
Here, $R_{\rm in}$ and $T_{rc}$ are the inner-radius and Coulomb-radius factors used in the analytical Coulomb-barrier expansion. Eqs. (\ref{eq37a}) and (\ref{eq37b}) show that $B_1$, $B_2$, and $B_3$ are proportional to $E(t)$, whereas $C_1$ is proportional to $E^2(t)$. Therefore, a positive electric field and a negative electric field produce opposite first-order contributions to $\Delta P$, while the field-squared part keeps the same sign. For a nearly symmetric monochromatic Gaussian pulse, the positive and negative half cycles consequently cancel each other strongly in the time average. In previous studies, several schemes have been proposed to reduce this cancellation. Short rectangular pulses with selected half cycles \cite{Mi_icu_2019}, elliptically polarized laser fields \cite{PhysRevC.99.044610}, and asymmetric chirped pulses \cite{PhysRevC.105.024312} can enhance the pulse-averaged effect by modifying the temporal or angular symmetry of the field. However, these strategies also have practical restrictions. The rectangular-pulse scheme requires the effective selection of only part of the oscillating field, which is difficult with current techniques. The elliptically polarized scheme mainly changes the angular projection between $\vec{E}(t)$ and $\vec{r}$, but it does not directly reshape the temporal frequency structure of the pulse. The chirped-pulse route can break the symmetry by detuning the compressor in a chirped-pulse-amplification system, but the useful chirp amount in a realistic laser pulse is limited. Our previous even-even calculation also showed that tuning the laser wavelength is more effective than tuning the pulse width for increasing $\Delta P_{\rm avg}$ \cite{CHENG2024138322}; nevertheless, the wavelength effect becomes especially evident only when the laser intensity is high enough or when shorter-wavelength facilities are available. It is therefore useful to search for a more universal scheme that works over different laser intensities, is experimentally accessible, and does not require increasing the fundamental peak intensity beyond the target laser condition.

\begin{figure}[!tbp]
\centering
\includegraphics[width=0.95\columnwidth]{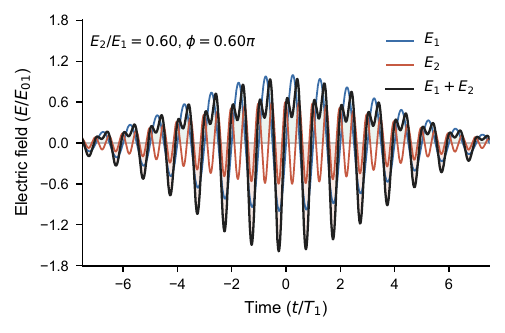}
\caption{Two-color laser waveform used to enhance the complete-pulse average response. The blue curve is the 800 nm fundamental component $E_1$, the red curve is the 400 nm second-harmonic component $E_2$, and the black curve is the total field $E_1+E_2$. The parameters are fixed at $E_2/E_1=0.60$ and $\phi=0.60\pi$.}
\label{fig5}
\end{figure}

To break the temporal symmetry of the laser electric field, we introduce a two-color Gaussian pulse by adding a second-harmonic component to the original 800 nm field. The electric field used in the present calculation is written as
\begin{equation}
\begin{aligned}
E_{\rm tc}(t)={}&E_{01}\exp\left(-\frac{c^2t^2}{x^2\lambda_1^2}\right)\\
&\times\left[\sin(\omega t)+\kappa\sin(2\omega t+\phi)\right],
\end{aligned}
\label{eq38}
\end{equation}
where $\lambda_1=800$ nm, $\lambda_2=400$ nm, $\omega=2\pi c/\lambda_1$, and $x=5$. The parameter $\kappa=E_{02}/E_{01}$ is the ratio between the electric-field amplitudes of the second-harmonic and fundamental components, and $\phi$ is their relative phase. In this work, the parameters are fixed as $\kappa=0.60$ and $\phi=0.60\pi$, which were selected from the parameter scan for the sensitive nucleus $^{236}$Np. Since the laser intensity is proportional to the square of the electric-field amplitude, this choice corresponds to $I_2/I_1=\kappa^2=0.36$. Thus, the second-harmonic field is weaker than the fundamental field. For the three fundamental peak intensities considered here, $I_1=10^{23}$, $10^{24}$, and $10^{25}\,\mathrm{W}/\mathrm{cm}^{2}$, the corresponding second-harmonic intensities are $3.6\times10^{22}$, $3.6\times10^{23}$, and $3.6\times10^{24}\,\mathrm{W}/\mathrm{cm}^{2}$, respectively.

The waveform generated by Eq. (\ref{eq38}) is shown in Fig. \ref{fig5}. The blue and red curves represent the fundamental and second-harmonic components, respectively, and the black curve represents their coherent superposition. Different from a symmetric monochromatic Gaussian pulse, the two-color pulse has unequal positive and negative half cycles. This half-cycle asymmetry is the key reason why two-color laser pulses have been widely used to control strong-field processes \cite{PhysRevA.100.033407, PhysRevLett.123.174801}. For the present $\alpha$ decay problem, the same symmetry-breaking mechanism is useful because it prevents the $E(t)$-linear terms in Eq. (\ref{eq37a}) from being completely cancelled in the pulse average. In other words, the two-color field does not change the nuclear structure itself, but changes the temporal weighting with which the external field samples the tunneling barrier.

For each odd-odd nucleus, we calculate the complete-pulse average penetration-probability variation for the monochromatic pulse and the two-color pulse. The pulse-averaged quantity is defined as
\begin{equation}
\Delta P_{\rm avg}=\frac{1}{2t_1}\int_{-t_1}^{t_1}\Delta P(t)dt,
\label{eq39}
\end{equation}
where $t_1$ is the half-width of the time interval used in the numerical pulse integration. The numerical values are listed in Table \ref{tab4}. The second and third columns give $\Delta P_{\rm avg}^{\rm m}$ and $\Delta P_{\rm avg}^{\rm tc}$ at $I_1=10^{23}\,\mathrm{W}/\mathrm{cm}^{2}$, respectively. The fourth and fifth columns give the corresponding quantities at $I_1=10^{24}\,\mathrm{W}/\mathrm{cm}^{2}$, and the last two columns give those at $I_1=10^{25}\,\mathrm{W}/\mathrm{cm}^{2}$. These values are several orders of magnitude smaller than the fixed-field $\Delta P^{23}$ values in Table \ref{tab1}, which directly reflects the cancellation over the complete laser pulse. Nevertheless, the comparison between adjacent monochromatic and two-color columns shows that the two-color pulse can systematically change the pulse-averaged response. For example, at $10^{24}\,\mathrm{W}/\mathrm{cm}^{2}$, $^{202}$At changes from $\Delta P_{\rm avg}^{\rm m}={3.269\times10^{-8}}$ to $\Delta P_{\rm avg}^{\rm tc}={4.331\times10^{-8}}$. At $10^{25}\,\mathrm{W}/\mathrm{cm}^{2}$, $^{216}$Fr changes from ${5.601\times10^{-8}}$ to ${8.467\times10^{-8}}$. This comparison indicates that the two-color pulse is not merely equivalent to increasing the laser intensity, but modifies the temporal symmetry of the driving field and makes the pulse-averaged tunneling response more favorable.

To quantify the gain produced by the two-color pulse, two relative indicators are introduced.
\begin{equation}
\eta_{\rm s}=100\frac{\Delta P_{\rm avg}^{\rm tc}-\Delta P_{\rm avg}^{\rm m}}
{|\Delta P_{\rm avg}^{\rm m}|},
\label{eq40}
\end{equation}
and
\begin{equation}
\eta_{\rm abs}=100\frac{|\Delta P_{\rm avg}^{\rm tc}|-|\Delta P_{\rm avg}^{\rm m}|}
{|\Delta P_{\rm avg}^{\rm m}|}.
\label{eq41}
\end{equation}
Here, the superscripts ``m'' and ``tc'' denote the monochromatic and two-color pulses, respectively. The signed quantity $\eta_{\rm s}$ describes whether the two-color field shifts $\Delta P_{\rm avg}$ toward the positive or negative direction. In contrast, $\eta_{\rm abs}$ measures the enhancement of the magnitude of the pulse-averaged laser effect. Since the experimental goal is to make the laser-induced change easier to observe, $\eta_{\rm abs}$ is the more direct indicator for judging whether the two-color field gives a stronger modification of the penetration probability. Both indicators are expressed in percent.

\begin{table*}[p]
\centering
\footnotesize
\setlength{\tabcolsep}{2pt}
\renewcommand{\arraystretch}{0.925}
\caption{Complete-pulse average penetration-probability variation of odd-odd nuclei under monochromatic and two-color laser pulses. The labels m and tc denote the monochromatic and two-color pulses, respectively; the fundamental peak intensity $I_1$ is given in parentheses in the column headings. All values are dimensionless.}
\label{tab4}
\begin{tabular*}{\textwidth}{@{\extracolsep{\fill}}lcccccc@{}}
\CPCtoprule
\rule[-0.8ex]{0pt}{3.0ex}Nucleus & m ($10^{23}$ W/cm$^2$) & tc ($10^{23}$ W/cm$^2$) & m ($10^{24}$ W/cm$^2$) & tc ($10^{24}$ W/cm$^2$) & m ($10^{25}$ W/cm$^2$) & tc ($10^{25}$ W/cm$^2$)\\
\CPCmidrule
$^{110}$I & {$3.811\times10^{-8}$} & {$4.043\times10^{-8}$} & {$5.878\times10^{-8}$} & {$6.595\times10^{-8}$} & {$1.405\times10^{-7}$} & {$1.925\times10^{-7}$}\\
$^{112}$I & {$-1.430\times10^{-8}$} & {$3.810\times10^{-9}$} & {$8.481\times10^{-8}$} & {$9.306\times10^{-8}$} & {$4.881\times10^{-7}$} & {$6.108\times10^{-7}$}\\
$^{114}$Cs & {$-2.023\times10^{-8}$} & {$-1.781\times10^{-8}$} & {$-8.658\times10^{-10}$} & {$8.124\times10^{-9}$} & {$2.241\times10^{-7}$} & {$3.094\times10^{-7}$}\\
$^{152}$Ho & {$-8.802\times10^{-8}$} & {$-8.929\times10^{-8}$} & {$-7.080\times10^{-8}$} & {$-6.032\times10^{-8}$} & {$2.069\times10^{-7}$} & {$3.229\times10^{-7}$}\\
$^{154}$Ho & {$-5.837\times10^{-8}$} & {$-5.334\times10^{-8}$} & {$-2.286\times10^{-8}$} & {$-1.609\times10^{-8}$} & {$3.732\times10^{-7}$} & {$6.066\times10^{-7}$}\\
$^{154}$Tm & {$-2.895\times10^{-8}$} & {$-3.017\times10^{-8}$} & {$-3.474\times10^{-8}$} & {$-3.592\times10^{-8}$} & {$7.917\times10^{-8}$} & {$1.533\times10^{-7}$}\\
$^{156}$Tm & {$4.566\times10^{-8}$} & {$4.572\times10^{-8}$} & {$8.620\times10^{-8}$} & {$9.922\times10^{-8}$} & {$4.697\times10^{-7}$} & {$5.902\times10^{-7}$}\\
$^{158}$Lu & {$9.193\times10^{-9}$} & {$1.183\times10^{-8}$} & {$3.440\times10^{-8}$} & {$4.632\times10^{-8}$} & {$2.620\times10^{-7}$} & {$3.719\times10^{-7}$}\\
$^{158}$Ta & {$-8.446\times10^{-9}$} & {$-7.004\times10^{-9}$} & {$2.988\times10^{-9}$} & {$7.982\times10^{-9}$} & {$1.090\times10^{-7}$} & {$1.490\times10^{-7}$}\\
$^{160}$Re & {$-2.285\times10^{-8}$} & {$-2.637\times10^{-8}$} & {$-1.930\times10^{-8}$} & {$-2.129\times10^{-8}$} & {$4.235\times10^{-8}$} & {$6.542\times10^{-8}$}\\
$^{162}$Ta & {$-3.549\times10^{-8}$} & {$-3.555\times10^{-8}$} & {$-6.239\times10^{-8}$} & {$-7.036\times10^{-8}$} & {$9.428\times10^{-8}$} & {$1.657\times10^{-7}$}\\
$^{162}$Re & {$2.539\times10^{-8}$} & {$2.336\times10^{-8}$} & {$4.226\times10^{-8}$} & {$5.055\times10^{-8}$} & {$1.583\times10^{-7}$} & {$1.884\times10^{-7}$}\\
$^{166}$Re & {$-4.330\times10^{-8}$} & {$-4.247\times10^{-8}$} & {$-3.185\times10^{-8}$} & {$-2.581\times10^{-8}$} & {$1.281\times10^{-7}$} & {$1.912\times10^{-7}$}\\
$^{166}$Ir & {$3.122\times10^{-8}$} & {$3.375\times10^{-8}$} & {$4.314\times10^{-8}$} & {$4.787\times10^{-8}$} & {$1.256\times10^{-7}$} & {$1.593\times10^{-7}$}\\
$^{168}$Re & {$-5.441\times10^{-8}$} & {$-5.832\times10^{-8}$} & {$-3.957\times10^{-8}$} & {$-3.468\times10^{-8}$} & {$2.038\times10^{-7}$} & {$2.996\times10^{-7}$}\\
$^{168}$Ir & {$-1.109\times10^{-8}$} & {$-1.196\times10^{-8}$} & {$-2.219\times10^{-9}$} & {$-1.556\times10^{-10}$} & {$8.728\times10^{-8}$} & {$1.241\times10^{-7}$}\\
$^{170}$Au & {$1.791\times10^{-8}$} & {$1.859\times10^{-8}$} & {$2.433\times10^{-8}$} & {$2.494\times10^{-8}$} & {$8.706\times10^{-8}$} & {$1.092\times10^{-7}$}\\
$^{172}$Ir & {$5.928\times10^{-8}$} & {$5.922\times10^{-8}$} & {$8.104\times10^{-8}$} & {$8.668\times10^{-8}$} & {$2.474\times10^{-7}$} & {$3.073\times10^{-7}$}\\
$^{174}$Ir & {$-4.708\times10^{-8}$} & {$-5.302\times10^{-8}$} & {$-3.845\times10^{-8}$} & {$-3.969\times10^{-8}$} & {$1.336\times10^{-7}$} & {$2.052\times10^{-7}$}\\
$^{176}$Au & {$4.092\times10^{-8}$} & {$4.423\times10^{-8}$} & {$7.043\times10^{-8}$} & {$7.363\times10^{-8}$} & {$1.932\times10^{-7}$} & {$2.355\times10^{-7}$}\\
$^{180}$Tl & {$-1.519\times10^{-8}$} & {$-1.397\times10^{-8}$} & {$-5.267\times10^{-9}$} & {$8.141\times10^{-11}$} & {$9.959\times10^{-8}$} & {$1.445\times10^{-7}$}\\
$^{186}$Bi & {$4.087\times10^{-9}$} & {$6.278\times10^{-9}$} & {$1.168\times10^{-8}$} & {$1.649\times10^{-8}$} & {$8.642\times10^{-8}$} & {$1.130\times10^{-7}$}\\
$^{190}$Bi & {$2.301\times10^{-9}$} & {$3.549\times10^{-9}$} & {$1.003\times10^{-8}$} & {$1.584\times10^{-8}$} & {$1.342\times10^{-7}$} & {$1.798\times10^{-7}$}\\
$^{192}$Bi & {$-4.026\times10^{-8}$} & {$-3.814\times10^{-8}$} & {$-2.136\times10^{-8}$} & {$-1.626\times10^{-8}$} & {$1.530\times10^{-7}$} & {$2.214\times10^{-7}$}\\
$^{192}$At & {$-3.337\times10^{-8}$} & {$-3.963\times10^{-8}$} & {$-4.432\times10^{-8}$} & {$-5.377\times10^{-8}$} & {$2.440\times10^{-8}$} & {$3.859\times10^{-8}$}\\
$^{194}$Bi & {$-8.952\times10^{-10}$} & {$-9.510\times10^{-10}$} & {$2.296\times10^{-8}$} & {$3.133\times10^{-8}$} & {$2.611\times10^{-7}$} & {$3.564\times10^{-7}$}\\
$^{200}$At & {$-2.443\times10^{-8}$} & {$-2.094\times10^{-8}$} & {$-6.637\times10^{-9}$} & {$8.749\times10^{-11}$} & {$1.681\times10^{-7}$} & {$2.383\times10^{-7}$}\\
$^{200}$Fr & {$-2.125\times10^{-8}$} & {$-1.952\times10^{-8}$} & {$-1.421\times10^{-8}$} & {$-1.010\times10^{-8}$} & {$9.265\times10^{-8}$} & {$1.373\times10^{-7}$}\\
$^{202}$At & {$3.542\times10^{-9}$} & {$5.922\times10^{-9}$} & {$3.269\times10^{-8}$} & {$4.331\times10^{-8}$} & {$2.291\times10^{-7}$} & {$3.208\times10^{-7}$}\\
$^{204}$At & {$5.146\times10^{-8}$} & {$5.184\times10^{-8}$} & {$8.441\times10^{-8}$} & {$9.409\times10^{-8}$} & {$3.583\times10^{-7}$} & {$4.747\times10^{-7}$}\\
$^{204}$Fr & {$2.338\times10^{-9}$} & {$2.009\times10^{-9}$} & {$1.672\times10^{-8}$} & {$2.105\times10^{-8}$} & {$1.502\times10^{-7}$} & {$2.015\times10^{-7}$}\\
$^{206}$At & {$-2.267\times10^{-8}$} & {$-2.498\times10^{-8}$} & {$2.405\times10^{-9}$} & {$1.375\times10^{-8}$} & {$3.083\times10^{-7}$} & {$4.305\times10^{-7}$}\\
$^{206}$Fr & {$-2.046\times10^{-8}$} & {$-2.236\times10^{-8}$} & {$-1.295\times10^{-8}$} & {$-9.613\times10^{-9}$} & {$1.233\times10^{-7}$} & {$1.916\times10^{-7}$}\\
$^{206}$Ac & {$2.967\times10^{-8}$} & {$3.185\times10^{-8}$} & {$4.126\times10^{-8}$} & {$4.663\times10^{-8}$} & {$1.355\times10^{-7}$} & {$1.810\times10^{-7}$}\\
$^{208}$At & {$2.953\times10^{-8}$} & {$3.031\times10^{-8}$} & {$6.486\times10^{-8}$} & {$8.204\times10^{-8}$} & {$4.209\times10^{-7}$} & {$5.629\times10^{-7}$}\\
$^{208}$Fr & {$3.705\times10^{-9}$} & {$2.858\times10^{-9}$} & {$2.224\times10^{-8}$} & {$2.759\times10^{-8}$} & {$2.046\times10^{-7}$} & {$2.736\times10^{-7}$}\\
$^{210}$At & {$6.709\times10^{-8}$} & {$7.173\times10^{-8}$} & {$1.102\times10^{-7}$} & {$1.271\times10^{-7}$} & {$5.164\times10^{-7}$} & {$6.750\times10^{-7}$}\\
$^{210}$Fr & {$7.866\times10^{-9}$} & {$6.879\times10^{-9}$} & {$3.036\times10^{-8}$} & {$3.579\times10^{-8}$} & {$2.308\times10^{-7}$} & {$3.107\times10^{-7}$}\\
$^{212}$Bi & {$4.889\times10^{-8}$} & {$4.972\times10^{-8}$} & {$8.468\times10^{-8}$} & {$9.501\times10^{-8}$} & {$3.838\times10^{-7}$} & {$4.972\times10^{-7}$}\\
$^{212}$At & {$1.510\times10^{-8}$} & {$1.532\times10^{-8}$} & {$2.855\times10^{-8}$} & {$3.269\times10^{-8}$} & {$1.474\times10^{-7}$} & {$1.951\times10^{-7}$}\\
$^{212}$Fr & {$-1.847\times10^{-8}$} & {$-2.579\times10^{-8}$} & {$-2.207\times10^{-8}$} & {$-1.462\times10^{-8}$} & {$9.252\times10^{-8}$} & {$1.918\times10^{-7}$}\\
$^{212}$Pa & {$-5.751\times10^{-8}$} & {$-5.718\times10^{-8}$} & {$-5.138\times10^{-8}$} & {$-4.802\times10^{-8}$} & {$2.383\times10^{-8}$} & {$6.761\times10^{-8}$}\\
$^{214}$Bi & {$1.014\times10^{-7}$} & {$1.007\times10^{-7}$} & {$1.650\times10^{-7}$} & {$1.687\times10^{-7}$} & {$6.060\times10^{-7}$} & {$7.887\times10^{-7}$}\\
$^{214}$At & {$-9.774\times10^{-9}$} & {$-1.074\times10^{-8}$} & {$-4.451\times10^{-9}$} & {$-3.260\times10^{-9}$} & {$6.657\times10^{-8}$} & {$9.695\times10^{-8}$}\\
$^{214}$Fr & {$-2.455\times10^{-8}$} & {$-2.627\times10^{-8}$} & {$-1.951\times10^{-8}$} & {$-1.840\times10^{-8}$} & {$6.342\times10^{-8}$} & {$9.776\times10^{-8}$}\\
$^{216}$At & {$4.117\times10^{-8}$} & {$4.331\times10^{-8}$} & {$5.658\times10^{-8}$} & {$6.319\times10^{-8}$} & {$1.742\times10^{-7}$} & {$2.200\times10^{-7}$}\\
$^{216}$Fr & {$-1.301\times10^{-8}$} & {$-1.148\times10^{-8}$} & {$-8.209\times10^{-9}$} & {$-5.016\times10^{-9}$} & {$5.601\times10^{-8}$} & {$8.467\times10^{-8}$}\\
$^{216}$Ac & {$6.948\times10^{-9}$} & {$-5.098\times10^{-9}$} & {$1.684\times10^{-8}$} & {$6.935\times10^{-10}$} & {$7.877\times10^{-8}$} & {$1.011\times10^{-7}$}\\
$^{218}$At & {$1.544\times10^{-8}$} & {$1.648\times10^{-8}$} & {$4.301\times10^{-8}$} & {$4.519\times10^{-8}$} & {$2.464\times10^{-7}$} & {$3.263\times10^{-7}$}\\
$^{218}$Fr & {$1.049\times10^{-8}$} & {$1.162\times10^{-8}$} & {$2.336\times10^{-8}$} & {$2.859\times10^{-8}$} & {$1.396\times10^{-7}$} & {$1.841\times10^{-7}$}\\
$^{218}$Ac & {$8.213\times10^{-9}$} & {$6.987\times10^{-9}$} & {$1.801\times10^{-8}$} & {$2.264\times10^{-8}$} & {$8.296\times10^{-8}$} & {$1.033\times10^{-7}$}\\
$^{220}$Fr & {$-6.350\times10^{-9}$} & {$-5.326\times10^{-9}$} & {$1.554\times10^{-8}$} & {$2.428\times10^{-8}$} & {$1.898\times10^{-7}$} & {$2.996\times10^{-7}$}\\
$^{220}$Ac & {$-9.283\times10^{-11}$} & {$-1.067\times10^{-9}$} & {$9.363\times10^{-9}$} & {$1.331\times10^{-8}$} & {$1.137\times10^{-7}$} & {$1.514\times10^{-7}$}\\
$^{220}$Pa & {$3.045\times10^{-8}$} & {$3.060\times10^{-8}$} & {$4.159\times10^{-8}$} & {$4.197\times10^{-8}$} & {$1.023\times10^{-7}$} & {$1.303\times10^{-7}$}\\
$^{222}$Ac & {$-7.074\times10^{-8}$} & {$-9.335\times10^{-8}$} & {$-1.771\times10^{-7}$} & {$-1.579\times10^{-7}$} & {$-5.884\times10^{-8}$} & {$5.786\times10^{-8}$}\\
$^{224}$Ac & {$5.551\times10^{-9}$} & {$5.766\times10^{-9}$} & {$3.929\times10^{-8}$} & {$1.059\times10^{-8}$} & {$1.911\times10^{-7}$} & {$3.669\times10^{-7}$}\\
$^{224}$Pa & {$-3.523\times10^{-9}$} & {$-1.913\times10^{-9}$} & {$-2.375\times10^{-8}$} & {$2.473\times10^{-8}$} & {$1.943\times10^{-7}$} & {$2.519\times10^{-7}$}\\
$^{226}$Ac & {$6.543\times10^{-8}$} & {$7.236\times10^{-8}$} & {$1.301\times10^{-7}$} & {$1.669\times10^{-7}$} & {$9.416\times10^{-7}$} & {$1.158\times10^{-6}$}\\
$^{228}$Pa & {$1.120\times10^{-8}$} & {$1.141\times10^{-8}$} & {$-3.033\times10^{-8}$} & {$-3.446\times10^{-8}$} & {$8.770\times10^{-8}$} & {$2.767\times10^{-7}$}\\
$^{230}$Pa & {$5.284\times10^{-10}$} & {$2.284\times10^{-8}$} & {$2.097\times10^{-7}$} & {$2.953\times10^{-7}$} & {$1.195\times10^{-6}$} & {$1.347\times10^{-6}$}\\
$^{236}$Np & {$4.383\times10^{-8}$} & {$4.546\times10^{-8}$} & {$1.830\times10^{-7}$} & {$2.414\times10^{-7}$} & {$1.239\times10^{-6}$} & {$1.681\times10^{-6}$}\\
$^{236}$Am & {$-3.183\times10^{-8}$} & {$-1.755\times10^{-8}$} & {$1.179\times10^{-8}$} & {$-8.900\times10^{-8}$} & {$-7.231\times10^{-7}$} & {$-6.729\times10^{-7}$}\\
$^{244}$Bk & {$3.565\times10^{-8}$} & {$3.856\times10^{-8}$} & {$2.807\times10^{-7}$} & {$3.396\times10^{-7}$} & {$9.643\times10^{-7}$} & {$1.049\times10^{-6}$}\\
$^{248}$Es & {$-4.513\times10^{-8}$} & {$-4.235\times10^{-8}$} & {$2.865\times10^{-8}$} & {$5.114\times10^{-8}$} & {$4.416\times10^{-7}$} & {$4.379\times10^{-7}$}\\
$^{258}$Md & {$2.501\times10^{-8}$} & {$2.685\times10^{-8}$} & {$5.805\times10^{-8}$} & {$6.244\times10^{-8}$} & {$3.401\times10^{-7}$} & {$4.675\times10^{-7}$}\\
$^{272}$Rg & {$1.716\times10^{-8}$} & {$1.742\times10^{-8}$} & {$-7.037\times10^{-8}$} & {$-5.483\times10^{-8}$} & {$-2.021\times10^{-7}$} & {$-2.101\times10^{-7}$}\\
\CPCbottomrule
\end{tabular*}
\end{table*}

\begin{figure*}[!t]
\centering
\includegraphics[width=0.9\textwidth]{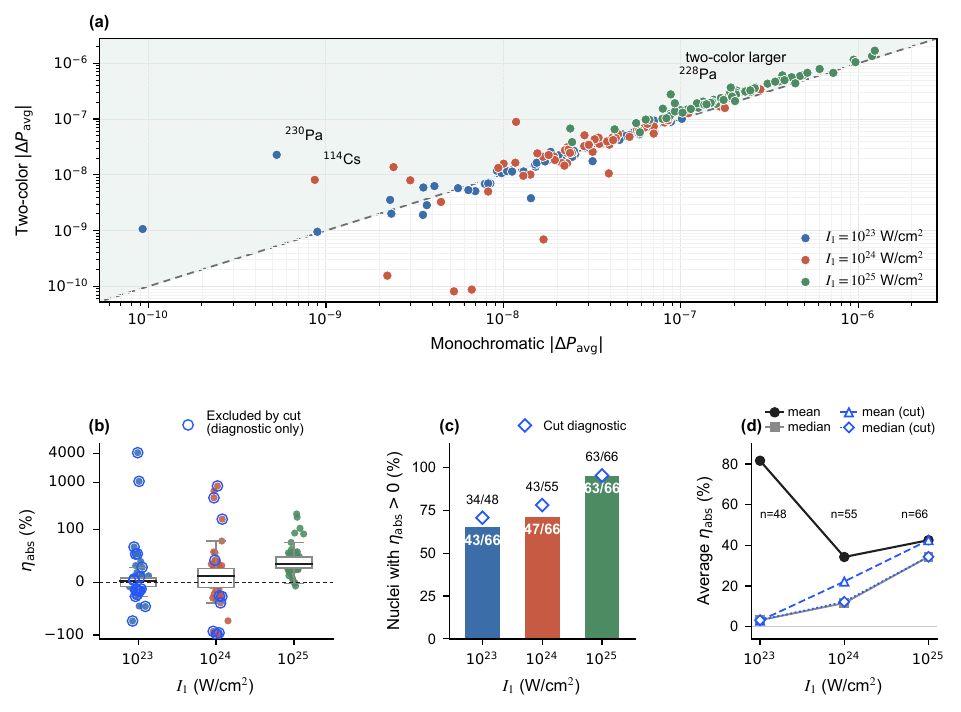}
\caption{Comparison between monochromatic and two-color complete-pulse average responses. (a) $|\Delta P_{\rm avg}^{\rm tc}|$ as a function of $|\Delta P_{\rm avg}^{\rm m}|$ for all selected odd-odd nuclei. The dashed line represents equality between the two pulses, and points above the line indicate that the two-color pulse gives a larger average effect. The nuclei with the largest $\eta_{\rm abs}$ at the three intensities are labeled. (b) Distribution of $\eta_{\rm abs}$ for the three fundamental peak intensities. All 66 nuclei are retained in each distribution. Blue open circles mark nuclei outside the subset defined by $|\Delta P_{\rm avg}^{\rm m}|>10^{-8}$ and $|\Delta P_{\rm avg}^{\rm tc}|>10^{-8}$. A symmetric logarithmic scale is used to display the low-intensity outliers. (c) Fraction of nuclei with $\eta_{\rm abs}>0$. The bars show the fractions for all 66 nuclei, and the blue diamonds show those for the selected subsets, which contain 48, 55, and 66 nuclei in order of increasing intensity. (d) Mean and median values of $\eta_{\rm abs}$. The black and gray curves show the full-sample mean and median, respectively; the blue curves show the corresponding values for the same subsets.}
\label{fig6}
\end{figure*}

\begin{figure*}[!t]
\centering
\includegraphics[width=0.9\textwidth]{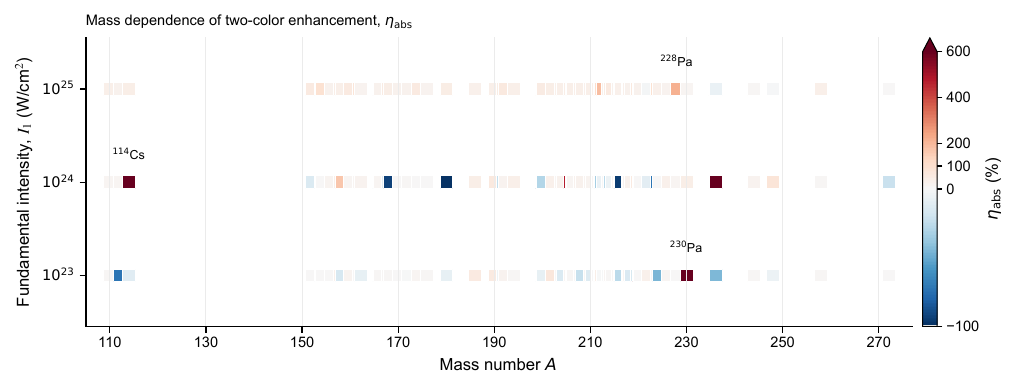}
\caption{Mass dependence of the two-color enhancement factor $\eta_{\rm abs}$, expressed in percent. The horizontal axis is the mass number of the parent nucleus, and the vertical axis gives the fundamental laser intensity $I_1$. Red colors denote nuclei for which the two-color pulse increases $|\Delta P_{\rm avg}|$, whereas blue colors denote a reduction of the magnitude. The most enhanced nucleus at each intensity is labeled.}
\label{fig7}
\end{figure*}

The global comparison between monochromatic and two-color pulses is shown in Fig. \ref{fig6}. Fig. \ref{fig6}(a) compares $|\Delta P_{\rm avg}^{\rm tc}|$ with $|\Delta P_{\rm avg}^{\rm m}|$ for all 66 odd-odd nuclei. Points above the diagonal line correspond to nuclei for which the two-color pulse produces a larger complete-pulse average effect. At $I_1=10^{23}\,\mathrm{W}/\mathrm{cm}^{2}$, the two-color advantage is still limited because the field-squared contribution is weak and the remaining first-order cancellation is important. Nevertheless, {43} of 66 nuclei already have $\eta_{\rm abs}>0$. When the fundamental intensity increases to $10^{24}\,\mathrm{W}/\mathrm{cm}^{2}$, the number of enhanced nuclei rises to {47}. At $10^{25}\,\mathrm{W}/\mathrm{cm}^{2}$, 63 of 66 nuclei show a larger $|\Delta P_{\rm avg}|$ under the two-color pulse. Figure \ref{fig6}(b) gives the distribution of $\eta_{\rm abs}$ at the three intensities. {The full-sample mean values of $\eta_{\rm abs}$ are $81.67\%$, $34.27\%$, and $42.55\%$, respectively; the low-intensity mean is strongly affected by small monochromatic denominators.} To make the global tendency clearer, Fig. \ref{fig6}(c) gives the fraction of nuclei for which $\eta_{\rm abs}$ is positive. This quantity does not describe the size of the largest enhancement, but measures how widely the two-color advantage appears in the selected odd-odd nuclei. The fraction increases from {$65.15\%$} at $10^{23}\,\mathrm{W}/\mathrm{cm}^{2}$ to {$71.21\%$} at $10^{24}\,\mathrm{W}/\mathrm{cm}^{2}$ and $95.45\%$ at $10^{25}\,\mathrm{W}/\mathrm{cm}^{2}$. Therefore, increasing the fundamental intensity not only amplifies several sensitive nuclei, but also changes the two-color pulse from a nucleus-dependent correction into a nearly global enhancement of the pulse-averaged tunneling response. Figure \ref{fig6}(d) further compares the mean and median values of $\eta_{\rm abs}$. The median value represents the typical nucleus and is less affected by a few strongly enhanced cases. {At $10^{23}\,\mathrm{W}/\mathrm{cm}^{2}$, the median is $3.17\%$ and the dual-threshold mean is $2.89\%$, indicating a weak typical enhancement; the much larger full-sample mean is caused by small monochromatic denominators, as discussed below.} At $10^{24}\,\mathrm{W}/\mathrm{cm}^{2}$, the mean value becomes much larger than the median value, which suggests that several nuclei with particularly favorable barrier sensitivity contribute strongly to the average enhancement. At $10^{25}\,\mathrm{W}/\mathrm{cm}^{2}$, the median value rises to {$34.41\%$} and approaches the mean value, showing that the enhancement is no longer dominated only by outlying nuclei. This behavior is consistent with the increasing role of the positive $E^2(t)$ contribution and the reduced cancellation of the $E(t)$-linear terms in the asymmetric two-color waveform.

\begingroup
A large percentage change need not imply a large absolute change in the pulse-averaged response. For $^{230}\mathrm{Pa}$ at $I_1=10^{23}\,\mathrm{W}/\mathrm{cm}^{2}$ in Table~\ref{tab4}, the monochromatic and two-color responses are approximately $5.28\times10^{-10}$ and $2.28\times10^{-8}$, respectively. Their absolute difference is $2.23\times10^{-8}$, whereas $\eta_{\rm abs}\approx4221.92\%$ because the monochromatic response is close to zero. For comparable absolute numerical errors, such small responses can have larger relative errors, motivating a statistical sensitivity check. To assess the influence of individual nuclei using the original unrounded data, we omit each nucleus in turn and recompute the statistics for the remaining 65 nuclei. At $I_1=10^{23}$, $10^{24}$, and $10^{25}\,\mathrm{W}/\mathrm{cm}^{2}$, the resulting mean values range from $17.98\%$ to $84.06\%$, $21.90\%$ to $36.32\%$, and $39.89\%$ to $43.31\%$, respectively. Figures~\ref{fig6}(b)--(d) also show a diagnostic selection with $|\Delta P_{\rm avg}^{\rm m}|>10^{-8}$ and $|\Delta P_{\rm avg}^{\rm tc}|>10^{-8}$, retaining 48, 55, and 66 nuclei. Requiring larger response magnitudes under both pulses reduces the potential relative amplification of comparable absolute numerical errors in either response; this threshold serves only as a statistical diagnostic. Their means are $2.89\%$, $22.19\%$, and $42.55\%$, and their medians are $3.17\%$, $12.20\%$, and $34.41\%$, respectively; 34, 43, and 63 retained nuclei have $\eta_{\rm abs}>0$. The original 66-nucleus statistics remain the main results. These checks support a weak typical enhancement at $10^{23}\,\mathrm{W}/\mathrm{cm}^{2}$ and enhancement for most nuclei at all three intensities, becoming nearly global at $10^{25}\,\mathrm{W}/\mathrm{cm}^{2}$. The low-intensity full-sample mean and the intermediate-intensity diagnostic mean are sensitive to individual small-response cases, so their magnitudes are interpreted together with the median and enhanced fraction.
\endgroup

The detailed mass dependence is displayed in Fig. \ref{fig7}. Each square represents one nucleus, and the color denotes $\eta_{\rm abs}$. At $10^{23}\,\mathrm{W}/\mathrm{cm}^{2}$, positive and negative enhancement regions coexist, which means that the fixed two-color phase does not improve every nucleus equally in the relatively weak field. {For $^{212}$Fr, log--log interpolation of the monotonically increasing monochromatic results gives $I_{\rm eff}/I_1\simeq12.84$ for the two-color response. For such sensitive nuclei, this enhancement corresponds to an effective gain that exceeds a simple one-order-of-magnitude increase in the monochromatic laser intensity.} At $10^{24}\,\mathrm{W}/\mathrm{cm}^{2}$, the enhanced region becomes broader, and the largest magnitude enhancement appears in {$^{114}$Cs with $\eta_{\rm abs}=838.35\%$. As discussed above, the exceptionally large percentage for $^{114}\mathrm{Cs}$ is amplified by its small monochromatic denominator. Nevertheless, at this intensity, the diagnostic subset retains positive mean and median enhancements of $22.19\%$ and $12.20\%$, respectively, and 43 of the 55 retained nuclei have $\eta_{\rm abs}>0$, indicating that the overall enhancement trend is not driven solely by this small-denominator case.}  At $10^{25}\,\mathrm{W}/\mathrm{cm}^{2}$, most nuclei are located in the positive-enhancement region, and the largest value appears in {$^{228}$Pa with $\eta_{\rm abs}=215.52\%$}. It should be emphasized that the nucleus giving the largest static or instantaneous laser response is not necessarily the same as the nucleus giving the largest two-color enhancement of the complete-pulse average. This is because the latter depends not only on the barrier sensitivity of the nucleus, but also on how the time-dependent field samples the $E(t)$-linear terms in Eq. (\ref{eq37a}) and the $E^2(t)$ contribution in Eq. (\ref{eq37b}).

Taken together, the direct numerical values in Table \ref{tab4} and the figure-level comparisons in Figs. \ref{fig6} and \ref{fig7} show that changing the temporal symmetry of the pulse provides an additional control variable for the complete-pulse average response. Therefore, the two-color laser pulse provides a practical and broadly applicable route for enhancing the average modification of the $\alpha$ decay penetration probability in odd-odd nuclei. 

\FloatBarrier

\section{Summary} 
\label{Sec.IV}
In summary, we have investigated the effect of ultra-intense laser fields on the $\alpha$ decay tunneling of deformed ground-state odd-odd nuclei within the deformed density-dependent cluster model and the WKB approximation. The calculated half-lives were first compared with the evaluated data, and the agreement supports the use of the adopted barrier description before introducing the external field. On this basis, we calculated the laser-induced variations in the penetration probability and half-life and found that the laser field changes the half-life mainly through the barrier penetration probability.

For deformed odd-odd $\alpha$ emitters, the laser response is not governed by $Q_{\alpha}$ alone. The laser-induced variation in the penetration probability separates into distinct branches associated with valence particle-hole configurations relative to the $Z=82$ and $N=126$ shell closures. The analysis of valence nucleon numbers, isospin asymmetry, and the empirical residual proton-neutron interaction further shows that structural correlations from the simultaneously unpaired proton and neutron are important. In contrast to even-even and odd-$A$ nuclei, the odd-odd parent nucleus most sensitive to the laser field is therefore not simply the one with the lowest $\alpha$ decay energy.

We also examined the complete-pulse average response. The oscillating symmetry of a monochromatic Gaussian pulse strongly cancels the average laser effect, whereas an optimized two-color field breaks this temporal symmetry. The two-color field enhances the pulse-averaged response for most of the studied nuclei, and highly sensitive nuclei show an effective gain comparable to increasing the laser intensity by more than one order of magnitude. These results identify structural determinants of laser-assisted $\alpha$ decay in odd-odd nuclei and provide guidance for selecting candidate nuclei and optimizing laser parameters in future experiments.


\end{document}